\documentclass[%
 amsmath,amssymb,
 aps,
 prx,
 twocolumn,
 longbibliography 
]{revtex4-2}

\usepackage{graphicx}
\usepackage{dcolumn}
\usepackage{bm}
\usepackage{xcolor}
\usepackage{tikz}
\usepackage{xfrac}
\usepackage{blindtext}
\usepackage{times}
\usepackage{multirow}
\usepackage{array}
\usepackage{chemformula}
\usepackage{braket}
\usepackage{tikz}
\newcommand{\tikzscale}{0.6}

\usepackage{color}
\usepackage[colorlinks=true, urlcolor=blue, linkcolor=blue, citecolor=blue, pdftex]{hyperref}

\usepackage[normalem]{ulem}

\definecolor{darkblue}{HTML}{004D6B}
\definecolor{darkred}{HTML}{8c1515}
\definecolor{darkgreen}{HTML}{006400}
\usepackage{hyperref}
\hypersetup{
	colorlinks=true,
	urlcolor=darkred,
	citecolor=darkblue,
	linkcolor=darkred,
	breaklinks
}

\usepackage[caption=false]{subfig}

\graphicspath{{figures/}}

\newcommand{\be}{\begin{equation}}
\newcommand{\ee}{\end{equation}}

\newcommand{\bea}{\begin{eqnarray}}
\newcommand{\eea}{\end{eqnarray}}
\newcommand{\beal}{\begin{align}}
\newcommand{\eeal}{\end{align}}

\renewcommand{\vec}[1]{\ensuremath{\mathbf{#1}}}

\newcommand{\plaquetteTFIM}{
\ket{
  \protect\tikz[baseline=-0.5ex, scale=\tikzscale]{
  \protect\draw (0, 0) -- (1.5ex, 1.5ex);
  \protect\draw (1.5ex, 1.5ex) -- (3.0ex, 0.0ex);
  \protect\draw (3.0ex, 0.0ex) -- (1.5ex, -1.5ex);
  \protect\draw (1.5ex, -1.5ex) -- (0, 0);
  }
}
}
\newcommand{\plaquetteDiag}{
\ket{
  \protect\tikz[baseline=-0.5ex, scale=\tikzscale]{
  \protect\draw (0, 0) -- (1.5ex, 1.5ex);
  \protect\draw (1.5ex, 1.5ex) -- (3.0ex, 0.0ex);
  \protect\draw (3.0ex, 0.0ex) -- (1.5ex, -1.5ex);
  \protect\draw (1.5ex, -1.5ex) -- (0, 0);
  \protect\draw (0, -1.5ex) -- (3.0ex, 1.5ex);
  }
}
}
\newcommand{\plaquetteAntiDiag}{
\ket{
  \protect\tikz[baseline=-0.5ex, scale=\tikzscale]{
  \protect\draw (0, 0) -- (1.5ex, 1.5ex);
  \protect\draw (1.5ex, 1.5ex) -- (3.0ex, 0.0ex);
  \protect\draw (3.0ex, 0.0ex) -- (1.5ex, -1.5ex);
  \protect\draw (1.5ex, -1.5ex) -- (0, 0);
  \protect\draw (0, 1.5ex) -- (3.0ex, -1.5ex);
  }
}
}
\newcommand{\plaquetteCross}{
\ket{
  \protect\tikz[baseline=-0.5ex, scale=\tikzscale]{
  \protect\draw (0, 0) -- (1.5ex, 1.5ex);
  \protect\draw (1.5ex, 1.5ex) -- (3.0ex, 0.0ex);
  \protect\draw (3.0ex, 0.0ex) -- (1.5ex, -1.5ex);
  \protect\draw (1.5ex, -1.5ex) -- (0, 0);
  \protect\draw (0, 1.5ex) -- (3.0ex, -1.5ex);
  \protect\draw (1.5ex, -1.5ex) -- (0, 0);
  \protect\draw (0, -1.5ex) -- (3.0ex, 1.5ex);
  }
}
}

\begin{document}

\title{Symmetry-deformed toric codes and the quantum dimer model}

\author{Jiaxin Qiao}
\author{Yoshito Watanabe}
\author{Simon Trebst}
\affiliation{Institute for Theoretical Physics, University of Cologne, 50937 Cologne, Germany}

\begin{abstract}
  Motivated by the recent introduction of a $U(1)$-symmetric toric code model,
  we investigate symmetry-based deformations of topological order by systematically deconstructing the Gauss-law-enforcing star terms 
  of the toric code (TC) Hamiltonian. 
  This ``term-dropping" protocol introduces global symmetries that go beyond the alternative framework of 
  ``ungauging" topological order in symmetry-deformed models and gives rise to models  such as the $U(1)$TC or $XY$TC.
  These models inherit (emergent) subsystem symmetries (from the original 1-form symmetry of the TC) that can give rise
  to (subextensive) ground-state degeneracies, which can still be organized by the eigenvalues of Wilson loop operators.
  However, we demonstrate that these models do {\it not} support topological or fracton order (as has been conjectured in the literature)
  due to the loss of (emergent) gauge symmetry. 
  An extreme deformation of the TC is the quantum dimer model (QDM), which we discuss along the family of symmetry-deformed models
  from the perspective of subsystem symmetries,  sublattice modulation, and quantum order-by-disorder mechanisms
  resulting in rich phase diagrams. 
  For the QDM, this allows us to identify an emergent SO(2) symmetry for what appears to be a gapless ground state (by numerical standards)
  that is unstable to the formation of a plaquette valence bond solid upon sublattice modulation. 
 \end{abstract}

\maketitle

\section{Introduction\\[-2mm]}
Symmetry is a unifying principle in the classification of quantum many-body phases.  
Beginning with the familiar case of \emph{spontaneous symmetry breaking} (SSB)---where different phases correspond to distinct patterns of broken global symmetry---one can extend the classification to \emph{symmetric} yet short-range entangled ground states known as \emph{symmetry-protected topological} (SPT) phases~\cite{Chen2013}. These phases are stable against any local perturbation that preserves the protecting symmetry. Beyond this symmetry-based paradigm lies \emph{topological quantum order} (TO), a long-range entangled phase supporting anyonic excitations and a ground-state degeneracy fixed by the system's topology~\cite{Wen1990, Wen1990b, Senthil2000}. Many TO phases are best viewed as emergent gauge theories with local gauge groups; crucially, this gauge structure does not itself have to be an explicit symmetry of the microscopic Hamiltonian, but can be an emergent symmetry at low temperatures.

A versatile tool to connect symmetry, topology, and entanglement is {\it gauging}---the procedure which converts a ``global'' symmetry to a ``local'' gauge symmetry~\cite{Levin2012,Haegeman2015,Vijay2016}. If the underlying (global) symmetry group is sufficiently “large,” one can perform \emph{partial gauging}, whereby the resulting system can host long-range entangled TO enriched by a residual global symmetry, yielding \emph{symmetry-enriched topological} order (SET)~\cite{Lu2016, Barkeshli2019}. Although gauging and its inverse, ungauging, naturally uncover a hierarchical structure of phases within a tractable group-theoretic framework, the interplay between global and local symmetries that \emph{cannot} be related by gauging remains a subtle open problem.

\begin{figure*}
  \centering
  \includegraphics[width=0.85\linewidth]{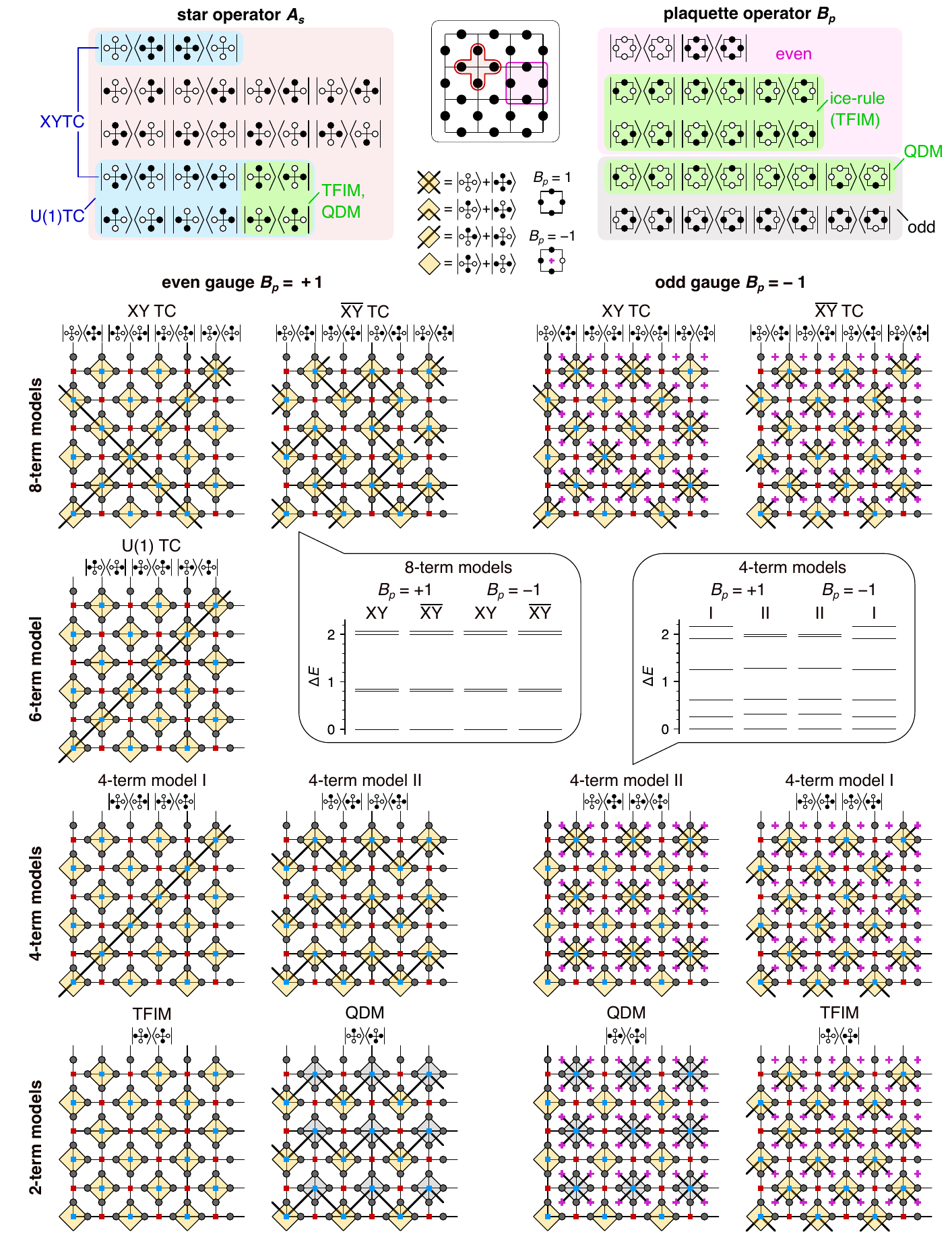}
  \caption{{\bf Schematic model overview} for the term-dropping protocol. 
  		Shown are instances of 8-term, 6-term, 4-term, and 2-term models that arise from restricting the 16 individual operators 
		(and their Hermitian conjugates) of the toric code's star terms shown in the upper left (orange shaded) panel.
		The plaquette term of the original toric code can be similarly represented (see the upper right, pink-shaded panel)
		where we distinguish terms enforcing $B_p=+1$ $(B_p=-1)$ plaquette eigenvalues corresponding to an even (odd) gauge theory. 
		For each model instance, we illustrate a representative example of a ground-state configuration in the exactly solvable limit
		of maximal sublattice modulation $(J_{s_2} = 0)$.
    		The black diagonal lines emanating from the plaquettes serve as a visual aid to identify the locations of fluxes, 
		as explained in detail in Fig.~\ref{fig:TGSD} and in the main text. 
		For the quantum dimer model (QDM), the gray plaquettes indicate those that do not gain energy from the star term
		(see the discussion in Sec.~\ref{sec:QDM}).
    }
  \label{fig:models}
\end{figure*}

In this work, we follow a different symmetry-deformation strategy to endow TO with global symmetry: instead of gauging or ungauging, we \emph{drop terms} from an exactly solvable Hamiltonian hosting a fixed-point TO phase. Specifically, we consider Kitaev's toric code (TC)~\cite{Kitaev2003} on a square lattice, which realizes the simplest $\mathbb{Z}_2$ gauge theory, and deconstruct the Gauss-law-enforcing star term, as illustrated in Fig.~\ref{fig:models}. We ask whether such a term-dropping protocol gives rise to a new hierarchical structure of symmetries, similar to the ungauging procedure.
We find that the answer is partially affirmative: 
Dropping terms can indeed lead to new subsystem or even global symmetries, 
with the most interesting example exhibiting a global $U(1)$ (instead of $\mathbb{Z}_2$) symmetry~\cite{Wu2023}.
But whether such a $U(1)$TC retains TO has remained an enigma---quantum Monte Carlo (QMC) simulations~\cite{Wu2023} indicated a ground-state degeneracy (GSD) of \emph{three} states, 
whose interpretation as a topological degeneracy points to the exhilarating possibility of non-Abelian TO (such as the Ising TQFT), but at the same time a sensitivity of this GSD to boundary conditions is counter-indicative of TO (with a 0 degree compactification reported~\cite{Wu2023} to result in a {\it two-fold} GSD).  


Motivated by this puzzle, we present a detailed analysis of the ground states of symmetry-deformed TC models, 
including the $U(1)$TC and a related $XY$TC~\cite{Vieweg2024} as important cases, and---in an extreme deformation---the quantum dimer model (QDM), which retains only a minimal number of terms, see Fig.~\ref{fig:models}. We show that symmetry plays a distinct role in all of these models. Crucially, we find that the acquisition of global symmetry comes at the expense of local (emergent) gauge symmetry, generically leading to a breakdown of TO. At the same time, an understanding of {\it subsystem symmetries} in these models leads to broad understanding of ground states and their degeneracies, including fresh insights into the QDM as the most extreme deformation of the TC.\\[-7mm]

\begin{figure*}
  \centering
  \includegraphics[width=\linewidth]{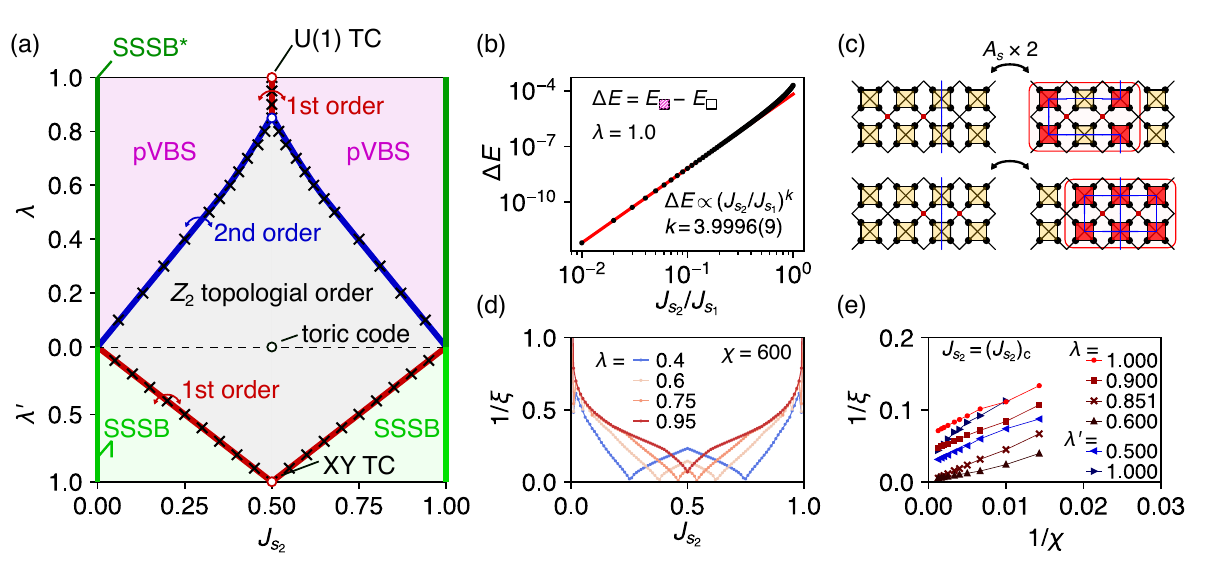}
  \caption{{\bf Phase diagram of symmetry-deformed toric codes.}
    (a) Phase diagram interpolating the $U(1)$TC and $XY$TC models to the TC model using parameters $\lambda$ and $\lambda'$, respectively, 
    	with a sublattice modulation of the star terms, $J_{s_2} = 1 - J_{s_1}$.
      The red/blue lines indicate first/second order transitions. The label ``pVBS" stands for plaquette valence bond solid, while ``SSSB" stands for spontaneous subsystem symmetry breaking phase. For the solvable limit of the $U(1)$TC the subsystem symmetry is emergent, indicated by ``SSSB*".
    (b) Energy difference between the full-line configuration and the pVBS configuration, $\Delta E = E_{\text{full-line}} - E_{\text{pVBS}}$.
    (c) Schematic illustration of the quantum order-by-disorder mechanism induced in 4th-order perturbation theory. 
    Energy costs are localized on the plaquettes marked in red.
    (d) The inverse correlation length $1/\xi$, which scales with the gap, along horizontal cuts in the upper half, 
    	indicating continuous phase transitions (with a gap closing) for $\lambda < 0.851$ and first-order transitions
	(with a finite gap) for $\lambda > 0.851$.
    (e)	 Scaling of the correlation length $\xi$ with bond dimension $\chi$ at phase boundaries.
    }
  \label{fig:phase_diagram_lambda_Js2}
\end{figure*}

\vskip -10mm

\section{Symmetry-deformed toric codes\\[-2mm]} 

Our starting point is the conventional TC Hamiltonian $H_{\text{TC}}~=~- \sum_{s}A_{s} - \sum_{p} B_{p}$, a sum of star and plaquette terms.  The latter, $B_{p} = \prod_{l \in \partial p} \sigma^z_l$, enforce a zero-flux constraint (which can be identified with the even sector of the underlying $Z_2$ gauge theory). Our attention lies on the star terms that, for the conventional TC, enforce a local Gauss-law, but will be modified for the  symmetry-deformed TC models by systematically removing operators from each star term
\begin{equation}
    \begin{split}
    A_{s} & =  \prod_{l \in s} \sigma^x_l = \prod_{l \in s}(\sigma^+_l + \sigma^-_l),\\
    & = \big(\ket{
      \tikz[baseline=-0.5ex, scale=\tikzscale]{
        \draw (1.0ex, -1.0ex) -- (1.0ex, 1.0ex);
        \draw (0, 0) -- (2ex, 0);
        \fill[white, draw=black] (0,0) circle (0.4ex);
        \fill[white, draw=black] (2ex,0) circle (0.4ex);
        \fill[white, draw=black] (1ex,1ex) circle (0.4ex);
        \fill[white, draw=black] (1ex,-1ex) circle (0.4ex);
      }
    }\!
    \bra{
      \tikz[baseline=-0.5ex, scale=\tikzscale]{
        \draw (1.0ex, -1.0ex) -- (1.0ex, 1.0ex);
        \draw (0, 0) -- (2ex, 0);
        \fill[black, draw=black] (0,0) circle (0.4ex);
        \fill[black, draw=black] (2ex,0) circle (0.4ex);
        \fill[black, draw=black] (1ex,1ex) circle (0.4ex);
        \fill[black, draw=black] (1ex,-1ex) circle (0.4ex);
      }
    }
    + 
    \ket{
      \tikz[baseline=-0.5ex, scale=\tikzscale]{
        \draw (1.0ex, -1.0ex) -- (1.0ex, 1.0ex);
        \draw (0, 0) -- (2ex, 0);
        \fill[black, draw=black] (0,0) circle (0.4ex);
        \fill[white, draw=black] (2ex,0) circle (0.4ex);
        \fill[white, draw=black] (1ex,1ex) circle (0.4ex);
        \fill[white, draw=black] (1ex,-1ex) circle (0.4ex);
      }
    }\!
    \bra{
      \tikz[baseline=-0.5ex, scale=\tikzscale]{
        \draw (1.0ex, -1.0ex) -- (1.0ex, 1.0ex);
        \draw (0, 0) -- (2ex, 0);
        \fill[white, draw=black] (0,0) circle (0.4ex);
        \fill[black, draw=black] (2ex,0) circle (0.4ex);
        \fill[black, draw=black] (1ex,1ex) circle (0.4ex);
        \fill[black, draw=black] (1ex,-1ex) circle (0.4ex);
      }
    }
    + 
    \ket{
      \tikz[baseline=-0.5ex, scale=\tikzscale]{
        \draw (1.0ex, -1.0ex) -- (1.0ex, 1.0ex);
        \draw (0, 0) -- (2ex, 0);
        \fill[white, draw=black] (0,0) circle (0.4ex);
        \fill[white, draw=black] (2ex,0) circle (0.4ex);
        \fill[black, draw=black] (1ex,1ex) circle (0.4ex);
        \fill[white, draw=black] (1ex,-1ex) circle (0.4ex);
      }
    }\!
    \bra{
      \tikz[baseline=-0.5ex, scale=\tikzscale]{
        \draw (1.0ex, -1.0ex) -- (1.0ex, 1.0ex);
        \draw (0, 0) -- (2ex, 0);
        \fill[black, draw=black] (0,0) circle (0.4ex);
        \fill[black, draw=black] (2ex,0) circle (0.4ex);
        \fill[white, draw=black] (1ex,1ex) circle (0.4ex);
        \fill[black, draw=black] (1ex,-1ex) circle (0.4ex);
      }
    }
    + 
    \ket{
      \tikz[baseline=-0.5ex, scale=\tikzscale]{
        \draw (1.0ex, -1.0ex) -- (1.0ex, 1.0ex);
        \draw (0, 0) -- (2ex, 0);
        \fill[white, draw=black] (0,0) circle (0.4ex);
        \fill[black, draw=black] (2ex,0) circle (0.4ex);
        \fill[white, draw=black] (1ex,1ex) circle (0.4ex);
        \fill[white, draw=black] (1ex,-1ex) circle (0.4ex);
      }
    }\!
    \bra{
      \tikz[baseline=-0.5ex, scale=\tikzscale]{
        \draw (1.0ex, -1.0ex) -- (1.0ex, 1.0ex);
        \draw (0, 0) -- (2ex, 0);
        \fill[black, draw=black] (0,0) circle (0.4ex);
        \fill[white, draw=black] (2ex,0) circle (0.4ex);
        \fill[black, draw=black] (1ex,1ex) circle (0.4ex);
        \fill[black, draw=black] (1ex,-1ex) circle (0.4ex);
      }
    } \\
    &
    + 
    \ket{
      \tikz[baseline=-0.5ex, scale=\tikzscale]{
        \draw (1.0ex, -1.0ex) -- (1.0ex, 1.0ex);
        \draw (0, 0) -- (2ex, 0);
        \fill[white, draw=black] (0,0) circle (0.4ex);
        \fill[white, draw=black] (2ex,0) circle (0.4ex);
        \fill[white, draw=black] (1ex,1ex) circle (0.4ex);
        \fill[black, draw=black] (1ex,-1ex) circle (0.4ex);
      }
    }\!
    \bra{
      \tikz[baseline=-0.5ex, scale=\tikzscale]{
        \draw (1.0ex, -1.0ex) -- (1.0ex, 1.0ex);
        \draw (0, 0) -- (2ex, 0);
        \fill[black, draw=black] (0,0) circle (0.4ex);
        \fill[black, draw=black] (2ex,0) circle (0.4ex);
        \fill[black, draw=black] (1ex,1ex) circle (0.4ex);
        \fill[white, draw=black] (1ex,-1ex) circle (0.4ex);
      }
    }
    +
    \ket{
      \tikz[baseline=-0.5ex, scale=\tikzscale]{
        \draw (1.0ex, -1.0ex) -- (1.0ex, 1.0ex);
        \draw (0, 0) -- (2ex, 0);
        \fill[black, draw=black] (0,0) circle (0.4ex);
        \fill[white, draw=black] (2ex,0) circle (0.4ex);
        \fill[black, draw=black] (1ex,1ex) circle (0.4ex);
        \fill[white, draw=black] (1ex,-1ex) circle (0.4ex);
      }
    }\!
    \bra{
      \tikz[baseline=-0.5ex, scale=\tikzscale]{
        \draw (1.0ex, -1.0ex) -- (1.0ex, 1.0ex);
        \draw (0, 0) -- (2ex, 0);
        \fill[white, draw=black] (0,0) circle (0.4ex);
        \fill[black, draw=black] (2ex,0) circle (0.4ex);
        \fill[white, draw=black] (1ex,1ex) circle (0.4ex);
        \fill[black, draw=black] (1ex,-1ex) circle (0.4ex);
      }
    }
    + 
    \ket{
      \tikz[baseline=-0.5ex, scale=\tikzscale]{
        \draw (1.0ex, -1.0ex) -- (1.0ex, 1.0ex);
        \draw (0, 0) -- (2ex, 0);
        \fill[black, draw=black] (0,0) circle (0.4ex);
        \fill[white, draw=black] (2ex,0) circle (0.4ex);
        \fill[white, draw=black] (1ex,1ex) circle (0.4ex);
        \fill[black, draw=black] (1ex,-1ex) circle (0.4ex);
      }
    }\!
    \bra{
      \tikz[baseline=-0.5ex, scale=\tikzscale]{
        \draw (1.0ex, -1.0ex) -- (1.0ex, 1.0ex);
        \draw (0, 0) -- (2ex, 0);
        \fill[white, draw=black] (0,0) circle (0.4ex);
        \fill[black, draw=black] (2ex,0) circle (0.4ex);
        \fill[black, draw=black] (1ex,1ex) circle (0.4ex);
        \fill[white, draw=black] (1ex,-1ex) circle (0.4ex);
      }
    }
    + 
    \ket{
      \tikz[baseline=-0.5ex, scale=\tikzscale]{
        \draw (1.0ex, -1.0ex) -- (1.0ex, 1.0ex);
        \draw (0, 0) -- (2ex, 0);
        \fill[black, draw=black] (0,0) circle (0.4ex);
        \fill[black, draw=black] (2ex,0) circle (0.4ex);
        \fill[white, draw=black] (1ex,1ex) circle (0.4ex);
        \fill[white, draw=black] (1ex,-1ex) circle (0.4ex);
      }
    }\!
    \bra{
      \tikz[baseline=-0.5ex, scale=\tikzscale]{
        \draw (1.0ex, -1.0ex) -- (1.0ex, 1.0ex);
        \draw (0, 0) -- (2ex, 0);
        \fill[white, draw=black] (0,0) circle (0.4ex);
        \fill[white, draw=black] (2ex,0) circle (0.4ex);
        \fill[black, draw=black] (1ex,1ex) circle (0.4ex);
        \fill[black, draw=black] (1ex,-1ex) circle (0.4ex);
      }
    } +\  \text{h.c.}
    \big),
    \end{split}
\end{equation}
where, in the second line, a pictorial representation is used for the $\sigma_z$-basis [black(white) for $\sigma_z = +1$($-1$)]. 
For example, retaining only the first and last three terms (and their Hermitian conjugates) yields the $XY$TC model \cite{Vieweg2024}, 
whose star term can also be expressed as $A_s = (\prod_{l \in s} \sigma^x_l + \prod_{l \in s} \sigma^y_l)/2$.
Subsequently removing the first term, leading to
$A_s =  \ket{
      \tikz[baseline=-0.5ex, scale=\tikzscale]{
        \draw (1.0ex, -1.0ex) -- (1.0ex, 1.0ex);
        \draw (0, 0) -- (2ex, 0);
        \fill[black, draw=black] (0,0) circle (0.4ex);
        \fill[white, draw=black] (2ex,0) circle (0.4ex);
        \fill[black, draw=black] (1ex,1ex) circle (0.4ex);
        \fill[white, draw=black] (1ex,-1ex) circle (0.4ex);
      }
    }\!
    \bra{
      \tikz[baseline=-0.5ex, scale=\tikzscale]{
        \draw (1.0ex, -1.0ex) -- (1.0ex, 1.0ex);
        \draw (0, 0) -- (2ex, 0);
        \fill[white, draw=black] (0,0) circle (0.4ex);
        \fill[black, draw=black] (2ex,0) circle (0.4ex);
        \fill[white, draw=black] (1ex,1ex) circle (0.4ex);
        \fill[black, draw=black] (1ex,-1ex) circle (0.4ex);
      }
    }
    + 
    \ket{
      \tikz[baseline=-0.5ex, scale=\tikzscale]{
        \draw (1.0ex, -1.0ex) -- (1.0ex, 1.0ex);
        \draw (0, 0) -- (2ex, 0);
        \fill[black, draw=black] (0,0) circle (0.4ex);
        \fill[white, draw=black] (2ex,0) circle (0.4ex);
        \fill[white, draw=black] (1ex,1ex) circle (0.4ex);
        \fill[black, draw=black] (1ex,-1ex) circle (0.4ex);
      }
    }\!
    \bra{
      \tikz[baseline=-0.5ex, scale=\tikzscale]{
        \draw (1.0ex, -1.0ex) -- (1.0ex, 1.0ex);
        \draw (0, 0) -- (2ex, 0);
        \fill[white, draw=black] (0,0) circle (0.4ex);
        \fill[black, draw=black] (2ex,0) circle (0.4ex);
        \fill[black, draw=black] (1ex,1ex) circle (0.4ex);
        \fill[white, draw=black] (1ex,-1ex) circle (0.4ex);
      }
    }
    + 
    \ket{
      \tikz[baseline=-0.5ex, scale=\tikzscale]{
        \draw (1.0ex, -1.0ex) -- (1.0ex, 1.0ex);
        \draw (0, 0) -- (2ex, 0);
        \fill[black, draw=black] (0,0) circle (0.4ex);
        \fill[black, draw=black] (2ex,0) circle (0.4ex);
        \fill[white, draw=black] (1ex,1ex) circle (0.4ex);
        \fill[white, draw=black] (1ex,-1ex) circle (0.4ex);
      }
    }\!
    \bra{
      \tikz[baseline=-0.5ex, scale=\tikzscale]{
        \draw (1.0ex, -1.0ex) -- (1.0ex, 1.0ex);
        \draw (0, 0) -- (2ex, 0);
        \fill[white, draw=black] (0,0) circle (0.4ex);
        \fill[white, draw=black] (2ex,0) circle (0.4ex);
        \fill[black, draw=black] (1ex,1ex) circle (0.4ex);
        \fill[black, draw=black] (1ex,-1ex) circle (0.4ex);
      }
    } +\  \text{h.c.}
$
, gives rise to a global $U(1)$ symmetry, resulting in the $U(1)$TC~\cite{Wu2023}.
One can further deform these models by keeping only four terms or even only two terms, where the model turns into an even gauge-sector equivalent of the well-known QDM.  A summary of these models is provided in Table \ref{tab:models} below.

\begin{figure}[b]
    \centering
    \includegraphics[width=0.9\linewidth]{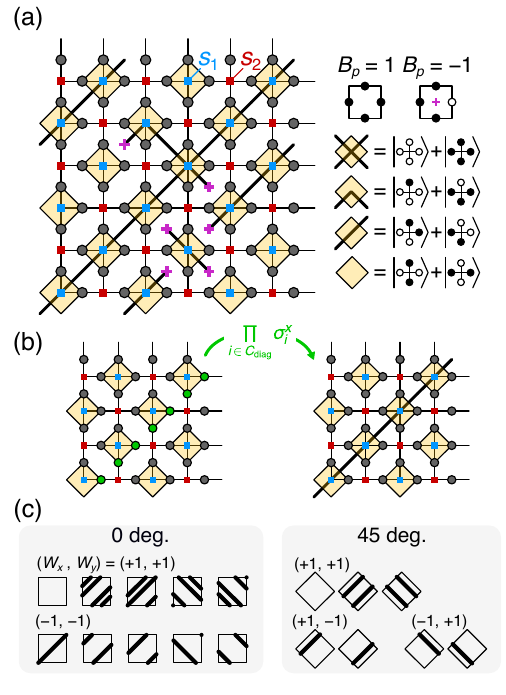}
    \caption{
    {\bf Graphical representation of ground-state configurations} in the exactly solvable limit $J_{s_2} = 0$.
    (a) The sites, on which the star terms act, are split into two sublattices, $s_1$ and $s_2$. 
    	Ground-state configurations can be constructed by assigning to each $s_1$ site a plaquette valence bond 
	favored by the corresponding $A_s$ term, 
	e.g., $\plaquetteTFIM$, $\plaquetteDiag$, or $\plaquetteAntiDiag$ in the case of the $U(1)$TC. 
	$B_p = -1$ plaquettes (indicated by a ``$+$") appear at the endpoints of (anti-)diagonal lines (see text for details). 
     (b) Ground-state configurations related by the $Z_2^{\text{sub}}$ symmetry can be sequentially generated by repeatedly applying the operator 
     	$\prod_{i \in C_{\text{diag}}} \sigma^x_i$, where $C_{\text{diag}}$ denotes the contour along the diagonal direction. 
     (c) Examples of the subextensive ground-state manifold of the $U(1)$TC, characterized by different compactification angles 
     	and classified by the eigenvalues of the operators $W_x$ and $W_y$.
    }
    \label{fig:TGSD}
\end{figure}

Any symmetry-deformed TC model satisfies $[B_p, B_{p'}] = [B_p, A_s] = 0$ and, on a torus geometry, the Hamiltonian commutes, 
for all models, with $W_x = \prod_{i \in C_x} \sigma^z_i$ and $W_y = \prod_{i \in C_y} \sigma^z_i$, 
where $C_x$ and $C_y$ denote contours along the periodic $x$ and $y$ directions, respectively.
The star terms, however, do {\it not} commute $[A_s, A_{s'}] \neq 0$ for neighboring $s$ sites---except in the undeformed TC case---indicating that the symmetry-deformed models are no longer integrable. 
To gain analytical insight we divide the star sites into two mutually staggered sublattices, $s_1$ and $s_2$: $- J_{s_1}\sum_{s_1} A_{s_1} - J_{s_2}\sum_{s_2} A_{s_2}$. When the star terms on the $s_2$ sublattice are turned off, the model decomposes into isolated clusters of four spins attached to the $s_1$ sites (subject to the $B_p$ constraints), thereby restoring integrability, see Fig.~\ref{fig:TGSD}(a). 
This provides, as we will discuss below, understanding of a broader GSD, in which a finite $J_{s_2}$ can then be treated as a perturbation.
Since the models are generally non-integrable, we also resort to numerical methods for a full understanding. Specifically, we employ iDMRG simulations~\cite{Schollwoeck2011}, for which the two-dimensional model is mapped onto a one-dimensional cylindrical geometry: the system is finite along the $y$ direction and infinite along the $x$ direction~\cite{Yan2011, Zaletel2013, Szasz2020, He2017}. To fully exploit the translational symmetry along the infinite direction, we carefully choose a compactification angle based on the expected symmetry-breaking pattern (see below and Appendix \ref{app:numerics} for further details).

\subsection{The $U(1)$ toric code}
We start our exploration of symmetry-deformed TC models by mapping out a phase dia\-gram (Fig.~\ref{fig:phase_diagram_lambda_Js2}) 
interpolating between the $U(1)$TC, the undeformed TC, and the $XY$TC Hamiltonians. For instance, in the upper half of the phase diagram we interpolate $H = (1 - \lambda)H_{\text{TC}} + \lambda H_{U(1)\text{TC}}$,
while also varying the relative strength of the two star terms $A_{s_1}$ and $A_{s_2}$ with $J_{s_2} = 1-J_{s_1}$. Similarly, we interpolate the TC to the $XY$TC (using a parameter $\lambda'$) in the lower half of the phase diagram.
The phase diagram shows a broad range of stability of TO around the TC fixed-point $(\lambda=\lambda'=0; J_{s_2}=1/2)$ to both
the symmetry deformation and sublattice modulation. 
The latter gives rise to ordered phases, a plaquette valence bond solid (pVBS) phase with a unique ground state and a spontaneous subsystem symmetry breaking (SSSB) phase. 
The $U(1)$TC and $XY$TC sit exactly at phase transitions between two symmetry-related sublattice ordering patterns. It is therefore instructive to start our further analysis from these ordered phases in their exactly solvable limits.

\subsubsection*{Exactly solvable limit} 
In the exactly solvable $J_{s_2} = 0$ limit of the $U(1)$TC the ground-state (GS) configurations for $B_p=+1$ plaquettes
can be graphically represented by a set of plaquette valence bonds centered at the $s_1$ sites
\[
\plaquetteTFIM =
\ket{
  \tikz[baseline=-0.5ex, scale=\tikzscale]{
    \draw (1.0ex, -1.0ex) -- (1.0ex, 1.0ex);
    \draw (0, 0) -- (2ex, 0);
    \fill[black, draw=black] (0,0) circle (0.4ex);
    \fill[black, draw=black] (2ex,0) circle (0.4ex);
    \fill[white, draw=black] (1ex,1ex) circle (0.4ex);
    \fill[white, draw=black] (1ex,-1ex) circle (0.4ex);
  }
} + 
\ket{
  \tikz[baseline=-0.5ex, scale=\tikzscale]{
    \draw (1.0ex, -1.0ex) -- (1.0ex, 1.0ex);
    \draw (0, 0) -- (2ex, 0);
    \fill[white, draw=black] (0,0) circle (0.4ex);
    \fill[white, draw=black] (2ex,0) circle (0.4ex);
    \fill[black, draw=black] (1ex,1ex) circle (0.4ex);
    \fill[black, draw=black] (1ex,-1ex) circle (0.4ex);
  }
}
\,,
\plaquetteDiag =
\ket{
  \tikz[baseline=-0.5ex, scale=\tikzscale]{
    \draw (1.0ex, -1.0ex) -- (1.0ex, 1.0ex);
    \draw (0, 0) -- (2ex, 0);
    \fill[black, draw=black] (0,0) circle (0.4ex);
    \fill[white, draw=black] (2ex,0) circle (0.4ex);
    \fill[white, draw=black] (1ex,1ex) circle (0.4ex);
    \fill[black, draw=black] (1ex,-1ex) circle (0.4ex);
  }
}+
\ket{
  \tikz[baseline=-0.5ex, scale=\tikzscale]{
    \draw (1.0ex, -1.0ex) -- (1.0ex, 1.0ex);
    \draw (0, 0) -- (2ex, 0);
    \fill[white, draw=black] (0,0) circle (0.4ex);
    \fill[black, draw=black] (2ex,0) circle (0.4ex);
    \fill[black, draw=black] (1ex,1ex) circle (0.4ex);
    \fill[white, draw=black] (1ex,-1ex) circle (0.4ex);
  }
}
\,, 
\plaquetteAntiDiag =
\ket{
  \tikz[baseline=-0.5ex, scale=\tikzscale]{
    \draw (1.0ex, -1.0ex) -- (1.0ex, 1.0ex);
    \draw (0, 0) -- (2ex, 0);
    \fill[black, draw=black] (0,0) circle (0.4ex);
    \fill[white, draw=black] (2ex,0) circle (0.4ex);
    \fill[black, draw=black] (1ex,1ex) circle (0.4ex);
    \fill[white, draw=black] (1ex,-1ex) circle (0.4ex);
  }
} + 
\ket{
  \tikz[baseline=-0.5ex, scale=\tikzscale]{
    \draw (1.0ex, -1.0ex) -- (1.0ex, 1.0ex);
    \draw (0, 0) -- (2ex, 0);
    \fill[white, draw=black] (0,0) circle (0.4ex);
    \fill[black, draw=black] (2ex,0) circle (0.4ex);
    \fill[white, draw=black] (1ex,1ex) circle (0.4ex);
    \fill[black, draw=black] (1ex,-1ex) circle (0.4ex);
  }
}
\,,
\]
with the pVBS configuration, $\prod_{s_1} \plaquetteTFIM_{s_1}$, being one of the ground states. In addition, each plaquette may have (anti) diagonal lines emanating from its center if the two spins on the corresponding edge have the same sign [Fig.~\ref{fig:TGSD}(a)]. If, in the multiplaquette state, these lines
have endpoints, then the corresponding plaquettes (at the end of the lines) will have a plaquette flux $B_p = -1$ and therefore describe an excited state.
If, on the other hand, the diagonal lines wrap around the torus, then we have another GS. 
The manifold of such states can be obtained by repeatedly applying the $\mathbb{Z}_2^{\text{sub}}$ symmetry operator, $\prod_{i \in C_{\text{diag}}} \sigma^x_i$, to the pVBS state [Fig.~\ref{fig:TGSD}(b)]. The same holds for ground states containing anti-diagonal line loops using $\prod_{i \in C_{\text{anti-diag}}}\sigma^x_i$ operators. 
This construction thereby leads to a $(2^{L/2 + 1} - 1)$-fold subextensive GSD in the solvable limit.

The GS manifold generated in this way exhibits, by  construction~\cite{Nussinov2015},  an emergent {\it subsystem symmetry} where the subextensive GSD is a direct consequence of SSB of this symmetry, indicated by SSSB* in Fig.~\ref{fig:phase_diagram_lambda_Js2}(a). In addition, GS can be organized by the eigenvalues of $W_x$ and $W_y$. Interestingly, the line representation, introduced as a visual tool to check the $B_p$ constraints, can also be used to determine the eigenvalues of $W_x$ and $W_y$, i.e., each time a loop crosses a line, the corresponding eigenvalue flips its sign. 
For even linear system size $L$, in the 0-degree compactification, this leads to two possibilities---$(W_x, W_y) = (\pm 1, \pm 1)$. Meanwhile, in the 45-degree compactification, three distinct sectors appear---$(W_x, W_y) = (+1, +1)$ and $(W_x, W_y) = (\pm 1, \mp 1)$ [see Fig.~\ref{fig:TGSD}(c)].

\subsubsection*{Quantum order-by-disorder}
Moving away from the solvable limit one can treat a finite $J_{s_2}$ in perturbation theory, which is expected to lift the classical GSD 
through a quantum order-by-disorder process. In finite-size systems, two types of quantum fluctuations can arise---global and local. 
Global fluctuations are induced by operators such as $\prod_{s_2 \in C'_{\text{diag}}} A_{s_2}$, or their anti-diagonal counterparts, which mediate hopping between different ground states (e.g.\ via inserting, removing, or shifting (anti-)diagonal lines). However, these processes are exponentially suppressed with the linear system size (and therefore invisible in our iDMRG simulations).
Local fluctuations, in contrast, act to locally dress the ground-state wave function.
Applying $A_{s_2}$ twice at the same $s_2$ site brings the system back to the same ground state, with intermediate energy costs localized on the four neighboring $s_1$ sites. \
This second-order process renormalizes the energy of all ground-state configurations equally and thus does not lift the degeneracy.

The situation changes at fourth order [Fig.~\ref{fig:phase_diagram_lambda_Js2}(b, c)]. When two neighboring $s_2$ terms are applied, the intermediate energy cost then depends on whether those two $s_2$ sites intersect with a (anti-)diagonal line. If they do, the excitation involves six $s_1$ sites; if not, the two central plaquettes incur no additional energy cost, as they are favored by the $U(1)$TC star term. As a result, the energy gain from this fourth-order process is maximized when the system contains no (anti-)diagonal lines, i.e., in the configuration $\prod_{s_1} \plaquetteTFIM_{s_1}$, corresponding to the pVBS state. This constitutes a quantum order-by-disorder mechanism \cite{Moessner2001} that selects the pVBS state as the unique ground state once $J_{s_2} \neq 0$. Numerically obtained energy splittings between the zero-line and full-line configurations confirm the fourth-order selection scenario [Fig.~\ref{fig:phase_diagram_lambda_Js2}(b)].
\subsubsection*{Ground state}

We find that the pVBS ground state selected at small but finite $J_{s_2} > 0$ persists up to the isotropic point $J_{s_1} = J_{s_2}$, where the $U(1)$TC model recovers full translational symmetry. This isotropic point can thus be interpreted either as a first-order transition point or as an SSB of translational symmetry, with the two possible pVBS states forming a two-fold GSD.
For any deviation $J_{s_1} \neq J_{s_2}$, translational symmetry is explicitly broken, and the ground state becomes unique.

In the presence of both global and local fluctuations, the {\it finite-size energy splitting} between the zero-line and single-line configurations is given by 
\[
	\Delta E(L) = \alpha \exp(-L/\xi) + \beta L \,
\]
with $L$ being the linear system size.
The first term arises from global fluctuations, while the second originates from local fluctuations and scales linearly in system size $L$. For small $L$, the first term dominates, leading to an apparent exponential suppression of the gap. As $L$ increases, however, the linear term becomes dominant and reopens the gap between topological sectors. Numerically, we estimate the prefactor $\beta$ to be rather small, approximately $1.1 \times 10^{-3}$, even at the isotropic point $J_{s_1} = J_{s_2}$. This strong numeral suppression of the finite-size effects of local fluctuations makes it a challenge to detect this splitting in any finite-size simulations.
For instance, to resolve this small energy splitting in QMC simulations, the temperature must be lower than the gap scale. Otherwise, the system appears to exhibit a renormalized classical ground-state manifold~\cite{Henry2014, Kato2015, Wang2025}. Consequently, rather than observing the true pVBS ground state, QMC might indicate an apparent ground-state degeneracy---two-fold for 0-degree and three-fold for 45-degree compactification---thus reconciling our results with previous QMC findings~\cite{Wu2023}.

\subsubsection*{Phase transitions}

Let us return to the phase diagram of Fig.~\ref{fig:phase_diagram_lambda_Js2}(a), which interpolates between the $U(1)$TC and the undeformed TC Hamiltonian and investigate how TO break downs at the various phase transitions. Starting from the $U(1)$TC point at $(\lambda=1, J_{s_1} = J_{s_2})$, and moving towards the TC along the $\lambda$-axis we observe a gap closing near $\lambda \approx 0.85$. This signals that the first-order transition at $J_{s_1} = J_{s_2}$ turns into a second-order phase transition, at which the two-fold degenerate pVBS phase meets the $\mathbb{Z}_2$ TO phase. In the extended parameter space with sublattice modulation $J_{s_1} \neq J_{s_2}$, this critical point turns out to be a {\it tricritical} point, from which two second-order transition lines branch out for $\lambda < 0.85$. We argue that these two transition lines are second-order transitions belonging to the universality class of the (2+1)-D Ising model---akin to the breakdown of TO in the toric code under a magnetic field~\cite{Trebst07toriccode,Tupitsyn2010,Dusuel2011,Troyer10topocrit}---for the following reasons:

Starting in the pVBS phase (e.g.\ for $J_{s_2} < J_{s_1}$), the $A_{s_2}$ term serves as a source of quantum fluctuations. Its action creates excitations on the four neighboring $s_1$ sites, which collectively form a domain wall, represented pictorially by the lines emanating from plaquettes introduced earlier. Repeated applications of $A_{s_2}$ extend these domain walls [cf.\ Fig.~\ref{fig:phase_diagram_lambda_Js2}(c)]. In addition to the explicit energy cost, illustrated by the red-colored plaquettes in the figure, there is also an implicit energy cost associated with the $\plaquetteDiag$ and $\plaquetteAntiDiag$ plaquettes. This cost arises from previously discussed fourth-order processes and from the reduced transition amplitude of the $A_{s_2}$ term (less than unity), akin to the physics of domain walls.

As $\lambda$ decreases, both the explicit energy cost and the suppression of the transition amplitude are reduced, lowering the overall `domain-wall energy'. This reduction drives the second-order transition observed for $\lambda < 0.85$, which can thus be interpreted as domain-wall condensation leading to the emergence of $\mathbb{Z}_2$ TO, analogous to the mechanism in the (2+1)-D Ising model~\cite{Kogut1979}. For $\lambda > 0.85$, quantum fluctuations are strongly suppressed, and the $\mathbb{Z}_2$ TO phase disappears entirely.

In terms of topology we can think of the pVBS phase as a topological trivial phase, in which the $m$ excitations of the TC are confined and the $e$ excitations condensed (with their condensation driving the phase transition from the TC into the pVBS phase). 
To see this, consider creating a pair of $m$ excitations ($B_p = -1$) in the pVBS phase and separating them along a diagonal direction. This process requires placing $\plaquetteDiag$ configurations along the path, with the $m$ excitations localized at the endpoints of the diagonal line. Since each $\plaquetteDiag$ configuration carries an energy cost, the $m$ excitations are confined. Equivalently, this reflects the condensation of $e$ excitations in the pVBS phase.
With this topological description in mind, note that our phase diagram indeed bears some resemblance to the TC phase diagram with a two-component magnetic field in the $XZ$-plane \cite{Tupitsyn2010,Dusuel2011} where---akin to the well-known Fradkin-Shenker phase diagram of the $\mathbb{Z}_2$ gauge theory \cite{FradkinShenker}---two lines of continuous Ising transitions meet in a tricritical point, out of which a first-order line emerges.
However, this latter tricritical point exhibits self-duality and the two continuous phase transitions are dual to one another. 
In contrast, no such duality connects the two continuous phase transition lines in our case. Both are transitions into staggered variants of the pVBS phase, each driven by the condensation of $e$ excitations and confinement of the $m$ vortices.

\subsection{The $XY$ toric code}

For the $XY$TC model, ground states can be constructed in a very similar manner to the $U(1)$TC: 
in the solvable limit $(J_{s_2} = 0)$ the expansion of the star operator to
\[
    A_{s} 
     = \ket{
      \tikz[baseline=-0.5ex, scale=\tikzscale]{
        \draw (1.0ex, -1.0ex) -- (1.0ex, 1.0ex);
        \draw (0, 0) -- (2ex, 0);
        \fill[white, draw=black] (0,0) circle (0.4ex);
        \fill[white, draw=black] (2ex,0) circle (0.4ex);
        \fill[white, draw=black] (1ex,1ex) circle (0.4ex);
        \fill[white, draw=black] (1ex,-1ex) circle (0.4ex);
      }
    }\!
    \bra{
      \tikz[baseline=-0.5ex, scale=\tikzscale]{
        \draw (1.0ex, -1.0ex) -- (1.0ex, 1.0ex);
        \draw (0, 0) -- (2ex, 0);
        \fill[black, draw=black] (0,0) circle (0.4ex);
        \fill[black, draw=black] (2ex,0) circle (0.4ex);
        \fill[black, draw=black] (1ex,1ex) circle (0.4ex);
        \fill[black, draw=black] (1ex,-1ex) circle (0.4ex);
      }
    }
    + 
    \ket{
      \tikz[baseline=-0.5ex, scale=\tikzscale]{
        \draw (1.0ex, -1.0ex) -- (1.0ex, 1.0ex);
        \draw (0, 0) -- (2ex, 0);
        \fill[black, draw=black] (0,0) circle (0.4ex);
        \fill[white, draw=black] (2ex,0) circle (0.4ex);
        \fill[black, draw=black] (1ex,1ex) circle (0.4ex);
        \fill[white, draw=black] (1ex,-1ex) circle (0.4ex);
      }
    }\!
    \bra{
      \tikz[baseline=-0.5ex, scale=\tikzscale]{
        \draw (1.0ex, -1.0ex) -- (1.0ex, 1.0ex);
        \draw (0, 0) -- (2ex, 0);
        \fill[white, draw=black] (0,0) circle (0.4ex);
        \fill[black, draw=black] (2ex,0) circle (0.4ex);
        \fill[white, draw=black] (1ex,1ex) circle (0.4ex);
        \fill[black, draw=black] (1ex,-1ex) circle (0.4ex);
      }
    }
    + 
    \ket{
      \tikz[baseline=-0.5ex, scale=\tikzscale]{
        \draw (1.0ex, -1.0ex) -- (1.0ex, 1.0ex);
        \draw (0, 0) -- (2ex, 0);
        \fill[black, draw=black] (0,0) circle (0.4ex);
        \fill[white, draw=black] (2ex,0) circle (0.4ex);
        \fill[white, draw=black] (1ex,1ex) circle (0.4ex);
        \fill[black, draw=black] (1ex,-1ex) circle (0.4ex);
      }
    }\!
    \bra{
      \tikz[baseline=-0.5ex, scale=\tikzscale]{
        \draw (1.0ex, -1.0ex) -- (1.0ex, 1.0ex);
        \draw (0, 0) -- (2ex, 0);
        \fill[white, draw=black] (0,0) circle (0.4ex);
        \fill[black, draw=black] (2ex,0) circle (0.4ex);
        \fill[black, draw=black] (1ex,1ex) circle (0.4ex);
        \fill[white, draw=black] (1ex,-1ex) circle (0.4ex);
      }
    }
    + 
    \ket{
      \tikz[baseline=-0.5ex, scale=\tikzscale]{
        \draw (1.0ex, -1.0ex) -- (1.0ex, 1.0ex);
        \draw (0, 0) -- (2ex, 0);
        \fill[black, draw=black] (0,0) circle (0.4ex);
        \fill[black, draw=black] (2ex,0) circle (0.4ex);
        \fill[white, draw=black] (1ex,1ex) circle (0.4ex);
        \fill[white, draw=black] (1ex,-1ex) circle (0.4ex);
      }
    }\!
    \bra{
      \tikz[baseline=-0.5ex, scale=\tikzscale]{
        \draw (1.0ex, -1.0ex) -- (1.0ex, 1.0ex);
        \draw (0, 0) -- (2ex, 0);
        \fill[white, draw=black] (0,0) circle (0.4ex);
        \fill[white, draw=black] (2ex,0) circle (0.4ex);
        \fill[black, draw=black] (1ex,1ex) circle (0.4ex);
        \fill[black, draw=black] (1ex,-1ex) circle (0.4ex);
      }
    } +\  \text{h.c.}
\]
inserts an additional plaquette state $\plaquetteCross = \ket{
  \tikz[baseline=-0.5ex, scale=\tikzscale]{
    \draw (1.0ex, -1.0ex) -- (1.0ex, 1.0ex);
    \draw (0, 0) -- (2ex, 0);
    \fill[white, draw=black] (0,0) circle (0.4ex);
    \fill[white, draw=black] (2ex,0) circle (0.4ex);
    \fill[white, draw=black] (1ex,1ex) circle (0.4ex);
    \fill[white, draw=black] (1ex,-1ex) circle (0.4ex);
  }
} + 
\ket{
  \tikz[baseline=-0.5ex, scale=\tikzscale]{
    \draw (1.0ex, -1.0ex) -- (1.0ex, 1.0ex);
    \draw (0, 0) -- (2ex, 0);
    \fill[black, draw=black] (0,0) circle (0.4ex);
    \fill[black, draw=black] (2ex,0) circle (0.4ex);
    \fill[black, draw=black] (1ex,1ex) circle (0.4ex);
    \fill[black, draw=black] (1ex,-1ex) circle (0.4ex);
  }
}$. The latter allows for the crossing of diagonal and anti-diagonal line loops (whose coexistence is forbidden in the $U(1)$TC). While the subsystem symmetry for the $U(1)$TC is an emergent symmetry, it becomes explicit for the $XY$TC \cite{Vieweg2024}. 
We note that this explicit subsystem symmetry $\mathbb{Z}_2^{\text{sub}}$ in the $XY$TC model descends directly from the original (magnetic) 1-form symmetry of the undeformed TC. 
The hallmark of a 1-form symmetry operator---the 't Hooft loop operator---is that it can be smoothly deformed, e.g.\ by multiplying it with the star terms, all of which commute with the Hamiltonian.  
If these star terms are removed (or modified as for the $XY$TC), this deformability is lost; the loop operator consequently splits into independent, rigid subsystem symmetries.
This symmetry setting then allows one to construct all $2^{L}$ distinct ground states in the solvable limit---each configuration is a realization of one of SSSB states.

\subsubsection*{Ground state}

For the $XY$TC model, increasing $J_{s_2} > 0$ away from the solvable limit we find that local fluctuations do {\it not} lift the GSD. As an illustrative example, consider the fourth-order perturbation process that selects the pVBS state in the $U(1)$TC model [Fig.~\ref{fig:phase_diagram_lambda_Js2}(c)]. In the $XY$TC case, the intermediate energy cost resides only on the four $s_1$ sites, since the $\plaquetteCross$ plaquettes---previously excited states in the $U(1)$TC model---are now energetically favored. Indeed, our numerical results, which are free from global fluctuations, show no energy splitting between the zero-line and full-line configurations, indicating that even higher-order perturbative processes fail to lift the GSD. This subextensive GSD persists up to the isotropic point $J_{s_1} = J_{s_2}$, where the system undergoes a first-order phase transition. This is also evidenced by the absence of a gap closing and a finite inverse correlation length at the transition [see Fig.~\ref{fig:phase_diagram_lambda_Js2}(e)]. 

The ground state of the $XY$TC model thus is described by a $2^{L+1}$ degenerate ground-state manifold, with each state being a realization of a spontaneous subsystem symmetry breaking (SSSB) state.

\subsubsection*{Phase transitions}

Previous work~\cite{Vieweg2024} on the $XY$TC model established a mapping between the $XY$TC model and two copies of the Xu-Moore model~\cite{Xu2004}. There, the transition between the $\mathbb{Z}_2$ TO phase (akin to the TC) and the SSSB phase (akin to the $XY$TC) corresponds to a {\it self-dual} point of the Xu-Moore model, which has been shown to undergo a first-order transition. Although our interpolation parameter $\lambda'$ between the $XY$TC and the TC model differs from the one used in Ref.~\cite{Vieweg2024}, the two can be brought into
one-to-one correspondence through a proper rescaling \footnote{
In Ref.~\cite{Vieweg2024}, an interpolation parameter $\phi$ is used to define the star term of the  $XY$TC as $A_s =  \cos{(\phi)}A^{(x)}_{s} + \sin{(\phi)}A^{(y)}_{s}$, 
where $A^{(\alpha)}_{s}$ is given by $\prod_{i \in s}\sigma^\alpha_i$ with $\alpha = x,y$. 
The mapping to our interpolation parameter $\lambda'$ can be expressed as $\phi = \arctan[\lambda'/(2-\lambda')]$. 
The self-dual point (at which the phase transition occurs) is then identified by the condition $J_{s_1} \cos(\phi) = J_{s_2} \sin(\phi)$. 
For $J_{s_1} = 1 - J_{s_2}$, this yields $J_{s_2} = 0.5 \lambda'$ for a given $\lambda'$, which agrees with our phase diagram shown in Fig.~\ref{fig:phase_diagram_lambda_Js2}(a).
}, implying that the phase transition lines in the lower half of our phase diagram, 
cf.\ Fig.~\ref{fig:phase_diagram_lambda_Js2}(a), are first-order transitions. This is indeed confirmed in our numerical simulations where the transition between the $\mathbb{Z}_2$ TO phase and the SSSB phase remains first-order (indicated by a finite gap/inverse correlation length) along the entire line.

\subsubsection*{Variations of the $XY${\rm TC} model}

The $XY$TC model, which comprises eight of the sixteen terms in the original TC, has a natural conjugate -- the $\overline{XY}$TC which is built from the remaining eight terms. 
In fact, this subset of eight terms is also a canonical choice since the terms, in their pictorial representation, are all symmetry-related to one another
\[A_s =
    \ket{
      \tikz[baseline=-0.5ex, scale=\tikzscale]{
        \draw (1.0ex, -1.0ex) -- (1.0ex, 1.0ex);
        \draw (0, 0) -- (2ex, 0);
        \fill[black, draw=black] (0,0) circle (0.4ex);
        \fill[white, draw=black] (2ex,0) circle (0.4ex);
        \fill[white, draw=black] (1ex,1ex) circle (0.4ex);
        \fill[white, draw=black] (1ex,-1ex) circle (0.4ex);
      }
    }\!
    \bra{
      \tikz[baseline=-0.5ex, scale=\tikzscale]{
        \draw (1.0ex, -1.0ex) -- (1.0ex, 1.0ex);
        \draw (0, 0) -- (2ex, 0);
        \fill[white, draw=black] (0,0) circle (0.4ex);
        \fill[black, draw=black] (2ex,0) circle (0.4ex);
        \fill[black, draw=black] (1ex,1ex) circle (0.4ex);
        \fill[black, draw=black] (1ex,-1ex) circle (0.4ex);
      }
    }
    + 
    \ket{
      \tikz[baseline=-0.5ex, scale=\tikzscale]{
        \draw (1.0ex, -1.0ex) -- (1.0ex, 1.0ex);
        \draw (0, 0) -- (2ex, 0);
        \fill[white, draw=black] (0,0) circle (0.4ex);
        \fill[white, draw=black] (2ex,0) circle (0.4ex);
        \fill[black, draw=black] (1ex,1ex) circle (0.4ex);
        \fill[white, draw=black] (1ex,-1ex) circle (0.4ex);
      }
    }\!
    \bra{
      \tikz[baseline=-0.5ex, scale=\tikzscale]{
        \draw (1.0ex, -1.0ex) -- (1.0ex, 1.0ex);
        \draw (0, 0) -- (2ex, 0);
        \fill[black, draw=black] (0,0) circle (0.4ex);
        \fill[black, draw=black] (2ex,0) circle (0.4ex);
        \fill[white, draw=black] (1ex,1ex) circle (0.4ex);
        \fill[black, draw=black] (1ex,-1ex) circle (0.4ex);
      }
    }
    + 
    \ket{
      \tikz[baseline=-0.5ex, scale=\tikzscale]{
        \draw (1.0ex, -1.0ex) -- (1.0ex, 1.0ex);
        \draw (0, 0) -- (2ex, 0);
        \fill[white, draw=black] (0,0) circle (0.4ex);
        \fill[black, draw=black] (2ex,0) circle (0.4ex);
        \fill[white, draw=black] (1ex,1ex) circle (0.4ex);
        \fill[white, draw=black] (1ex,-1ex) circle (0.4ex);
      }
    }\!
    \bra{
      \tikz[baseline=-0.5ex, scale=\tikzscale]{
        \draw (1.0ex, -1.0ex) -- (1.0ex, 1.0ex);
        \draw (0, 0) -- (2ex, 0);
        \fill[black, draw=black] (0,0) circle (0.4ex);
        \fill[white, draw=black] (2ex,0) circle (0.4ex);
        \fill[black, draw=black] (1ex,1ex) circle (0.4ex);
        \fill[black, draw=black] (1ex,-1ex) circle (0.4ex);
      }
    } 
    + 
    \ket{
      \tikz[baseline=-0.5ex, scale=\tikzscale]{
        \draw (1.0ex, -1.0ex) -- (1.0ex, 1.0ex);
        \draw (0, 0) -- (2ex, 0);
        \fill[white, draw=black] (0,0) circle (0.4ex);
        \fill[white, draw=black] (2ex,0) circle (0.4ex);
        \fill[white, draw=black] (1ex,1ex) circle (0.4ex);
        \fill[black, draw=black] (1ex,-1ex) circle (0.4ex);
      }
    }\!
    \bra{
      \tikz[baseline=-0.5ex, scale=\tikzscale]{
        \draw (1.0ex, -1.0ex) -- (1.0ex, 1.0ex);
        \draw (0, 0) -- (2ex, 0);
        \fill[black, draw=black] (0,0) circle (0.4ex);
        \fill[black, draw=black] (2ex,0) circle (0.4ex);
        \fill[black, draw=black] (1ex,1ex) circle (0.4ex);
        \fill[white, draw=black] (1ex,-1ex) circle (0.4ex);
      }
    } + \text{h.c.} 
\]
Notably, the $\overline{XY}$TC and the previously discussed $XY$TC are {\it isospectral}, i.e.\ their energy spectra are in one-to-one correspondence. This even holds when flipping the gauge constraint from 
$B_p = +1$ (even gauge) to $B_p = -1$ (odd gauge), as illustrated in the inset of Fig.~\ref{fig:models}. As a consequence, we can construct the ground-state manifold of the $\overline{XY}$TC in a manner analogous to what we have discussed above. Figure~\ref{fig:models} shows an example GS configuration of the symmetry-deformed  $\overline{XY}$TC  in the exactly solvable limit ($J_{s_2} = 0$), from which other ground state configurations can be systematically generated by applying the symmetry operator $\prod_{i \in C_{\text{(anti-)diag}}} \sigma^x_i$. 
The  $\overline{XY}$TC in the isotropic limit (without sublattice modulation) then also sits at a first-order transition where the two ground-state manifolds coexist, giving rise to a GSD of $2^{L+1}$ states, 
with each state being a realization of a spontaneous subsystem symmetry breaking (SSSB) state.

Interestingly, there are no other notable eight-term models beyond these $XY$TC models (and their conjugates). Every other subset of eight terms will reduce to an equivalent 6-, 4-, or 2-term model (with the remaining terms only renormalize states without causing qualitative changes), which we will discuss below.

\subsection{Hierarchy of models}
Our ``term-dropping" protocol has, so far, taken us from the the original, undeformed TC model where the star term is expressed in 16 terms
to its symmetry-deformed descendants, the $XY$TC and $U(1)$TC with $B_p = +1$ constraint, containing 8 and 6 terms, respectively. 

The full hierarchical structure of symmetry-deformed TC models is revealed by further removing terms, as summarized in Table~\ref{tab:models}. 
We focus on the case with uniform $B_p = 1$ or $B_p = -1$ for all $p$. The models that include all 16 possible terms correspond to the odd- and even-gauge TC models, associated with $B_p = 1$ and $B_p = -1$, respectively.

\begin{table}[t]
    \begin{tabular}{|l|c|c|c|}
    	\hline
    	\multirow{2}{*}{\bf model}		& \multirow{2}{*}{$\,$ terms $\,$} 	& $\quad$ subsystem $\quad$				& $\quad$ ground-state $\quad$			\\
    	 						& 	 				&  symmetry				&  degeneracy			\\
	\hline\hline	
    	toric code 			& 16		& 1-form					&	4-fold (topo.)				\\
	$XY$ toric code 	& 8		& explicit (rigid)					&	$2^{L+1}$ \\
	$U(1)$ toric code	& 6		& emergent$^\dagger$ (rigid)		&	2-fold (pVBS)					\\
	\hline\hline
	4-term model I$\quad$	& 4		& explicit (rigid)					&	$2^{L/2+1}$-fold					\\
	4-term model II		& 4		& ---						&	4-fold (pVBS)					\\
	\hline\hline
	TFIM$^*$			& 2		& ---						&	2-fold (pVBS)					\\
	QDM$^*$			& 2		& ---						&	critical						\\
	\hline
    \end{tabular}	
    \caption{{\bf Summary of the models discussed in this work. }
    		The ground-state degeneracy at $J_{s_1} = J_{s_2}$ is calculated for an $L \times L$ system with periodic boundary conditions.
		The asterisk ($^*$) for the transverse field Ising model (TFIM) and the quantum dimer model (QDM) indicates that 
		these models are defined in a modified gauge sector, but remain in one-to-one correspondence with their original definition.
		The $\dagger$ indicates that the subsystem symmetry is emergent in the limit  $(J_{s_2} = 0)$.}
    \label{tab:models}
\end{table}

\subsubsection*{4-term models}
One can identify two distinct types of models with just  4-term: 
(i) one with explicit subsystem symmetry (which we denote as `4-term~model~I'), whose GS realizes a SSSB phase, 
and (ii) another one without it (termed `4-term~model~II'), whose GS forms a 4-fold degenerate pVBS.

A concrete example of model I is defined by
\[A_s =
    \ket{
      \tikz[baseline=-0.5ex, scale=\tikzscale]{
        \draw (1.0ex, -1.0ex) -- (1.0ex, 1.0ex);
        \draw (0, 0) -- (2ex, 0);
        \fill[black, draw=black] (0,0) circle (0.4ex);
        \fill[white, draw=black] (2ex,0) circle (0.4ex);
        \fill[white, draw=black] (1ex,1ex) circle (0.4ex);
        \fill[black, draw=black] (1ex,-1ex) circle (0.4ex);
      }
    }\!
    \bra{
      \tikz[baseline=-0.5ex, scale=\tikzscale]{
        \draw (1.0ex, -1.0ex) -- (1.0ex, 1.0ex);
        \draw (0, 0) -- (2ex, 0);
        \fill[white, draw=black] (0,0) circle (0.4ex);
        \fill[black, draw=black] (2ex,0) circle (0.4ex);
        \fill[black, draw=black] (1ex,1ex) circle (0.4ex);
        \fill[white, draw=black] (1ex,-1ex) circle (0.4ex);
      }
    }
    + 
    \ket{
      \tikz[baseline=-0.5ex, scale=\tikzscale]{
        \draw (1.0ex, -1.0ex) -- (1.0ex, 1.0ex);
        \draw (0, 0) -- (2ex, 0);
        \fill[black, draw=black] (0,0) circle (0.4ex);
        \fill[black, draw=black] (2ex,0) circle (0.4ex);
        \fill[white, draw=black] (1ex,1ex) circle (0.4ex);
        \fill[white, draw=black] (1ex,-1ex) circle (0.4ex);
      }
    }\!
    \bra{
      \tikz[baseline=-0.5ex, scale=\tikzscale]{
        \draw (1.0ex, -1.0ex) -- (1.0ex, 1.0ex);
        \draw (0, 0) -- (2ex, 0);
        \fill[white, draw=black] (0,0) circle (0.4ex);
        \fill[white, draw=black] (2ex,0) circle (0.4ex);
        \fill[black, draw=black] (1ex,1ex) circle (0.4ex);
        \fill[black, draw=black] (1ex,-1ex) circle (0.4ex);
      }
    } +\  \text{h.c.} 
\]
The operator $\prod_{i \in C_{\text{diag}}} \sigma^x_i$ exchanges the two terms in $A_s$, indicating that the model possesses a diagonal subsystem $\mathbb{Z}_2^{\text{sub}}$ symmetry. In contrast to the $XY$TC model, however, $\prod_{i \in C_{\text{anti-diag}}} \sigma^x_i$ is not a symmetry of the Hamiltonian. At $J_{s_2} = 0$, the GS manifold is generated by applying the subsystem symmetry operator to a pVBS configuration $\prod_{s_1} \plaquetteTFIM_{s_1}$, resulting in a $2^{L/2}$-fold degeneracy characteristic of SSSB. Quantum fluctuations at finite $J_{s_2}$ do not lift this degeneracy, and the symmetric point $J_{s_1} = J_{s_2}$ (without sublattice modulation)  corresponds to a first-order transition between two distinct SSSB phases. 

Any set of four star terms that are related by $\prod_{i \in C_{\text{(anti-)diag}}} \sigma^x_i$ defines an equivalent model, provided the energetically favored plaquettes can be arranged to satisfy the $B_p$ constraint. A further example in the $B_p = +1$ sector is shown in Fig.~\ref{fig:models}, where equivalence is confirmed by isospectral ED results at $J_{s_1} = J_{s_2}$.

An example of model II is defined by 
\[ 
A_s =
    \ket{
      \tikz[baseline=-0.5ex, scale=\tikzscale]{
        \draw (1.0ex, -1.0ex) -- (1.0ex, 1.0ex);
        \draw (0, 0) -- (2ex, 0);
        \fill[white, draw=black] (0,0) circle (0.4ex);
        \fill[white, draw=black] (2ex,0) circle (0.4ex);
        \fill[black, draw=black] (1ex,1ex) circle (0.4ex);
        \fill[white, draw=black] (1ex,-1ex) circle (0.4ex);
      }
    }\!
    \bra{
      \tikz[baseline=-0.5ex, scale=\tikzscale]{
        \draw (1.0ex, -1.0ex) -- (1.0ex, 1.0ex);
        \draw (0, 0) -- (2ex, 0);
        \fill[black, draw=black] (0,0) circle (0.4ex);
        \fill[black, draw=black] (2ex,0) circle (0.4ex);
        \fill[white, draw=black] (1ex,1ex) circle (0.4ex);
        \fill[black, draw=black] (1ex,-1ex) circle (0.4ex);
      }
    }
    + 
    \ket{
      \tikz[baseline=-0.5ex, scale=\tikzscale]{
        \draw (1.0ex, -1.0ex) -- (1.0ex, 1.0ex);
        \draw (0, 0) -- (2ex, 0);
        \fill[white, draw=black] (0,0) circle (0.4ex);
        \fill[white, draw=black] (2ex,0) circle (0.4ex);
        \fill[white, draw=black] (1ex,1ex) circle (0.4ex);
        \fill[black, draw=black] (1ex,-1ex) circle (0.4ex);
      }
    }\!
    \bra{
      \tikz[baseline=-0.5ex, scale=\tikzscale]{
        \draw (1.0ex, -1.0ex) -- (1.0ex, 1.0ex);
        \draw (0, 0) -- (2ex, 0);
        \fill[black, draw=black] (0,0) circle (0.4ex);
        \fill[black, draw=black] (2ex,0) circle (0.4ex);
        \fill[black, draw=black] (1ex,1ex) circle (0.4ex);
        \fill[white, draw=black] (1ex,-1ex) circle (0.4ex);
      }
    } + \text{h.c.}
\]
with $B_p = +1$. In this case, the model does not exhibit a subsystem symmetry, as the two terms in $A_s$ are not related by $\prod_{i \in C_{\text{(anti-)diag}}} \sigma^x_i$. One may interpret the terms in $A_s$ as favoring two complementary types of plaquette configurations. To satisfy the $B_p = +1$ constraint globally, both types must be present; using only one type would violate the constraint (see Fig.~\ref{fig:models}). As a result, the GS exhibits a 2-fold degeneracy in the exactly solvable limit, which increases to 4-fold at the symmetric point $J_{s_1} = J_{s_2}$. An equivalent model can also be realized in the $B_p = -1$ sector; an example is shown in Fig.~\ref{fig:models}.

\subsubsection*{2-term models : TFIM and QDM}

\begin{figure}[b]
  \centering
  \includegraphics[width=1.0\linewidth]{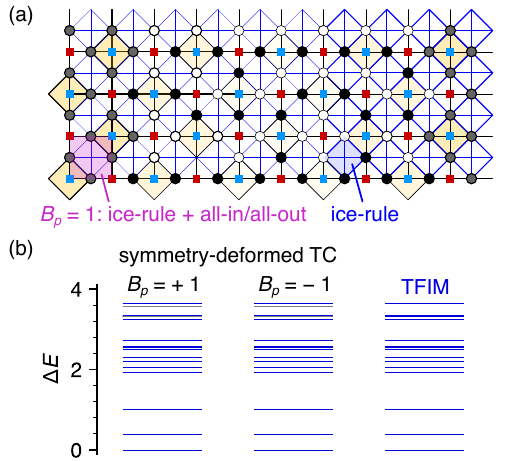}
  \caption{{\bf Equivalence between TFIM and symmetry-deformed TC models.}
  (a) Schematics of a spin configuration for symmetry-deformed TC/TFIM.
  (b) Energy spectra of two variants of the TFIM-type two-term symmetry-deformed TC model with uniform $B_p = +1$ and $B_p = -1$ constraints and the TFIM , obtained via ED for a system of size $N = 32$. While all three models share the same low-energy spectra (indicated by the blue lines), the symmetry-deformed TC models also exhibit additional states that are dynamically disconnected from the TFIM subspace.}
  \label{fig:TFIM}
\end{figure}

The most reduced cases are 2-term models, obtained by selecting two terms from the above 4-term models. Their GS correspond either to that of the transverse-field antiferromagnetic Ising model (TFIM) on the checkerboard lattice (2-fold pVBS)~\cite{Henry2014} or to the QDM, which is the only model in the hierarchy that appears to have a quantum critical ground state, as discussed in the next section.

The checkerboard lattice, also known as the 2D pyrochlore lattice, consists of a corner-sharing network of tetrahedra [shown in blue in Fig.~\ref{fig:TFIM}(a)]. In the absence of the transverse field, the TFIM reduces to a frustrated classical Ising model, where each tetrahedron satisfies the two-in-two-out ``ice-rule'' constraint, leading to an extensive ground-state degeneracy. The transverse field introduces quantum fluctuations, whose leading nontrivial process is
$
  \ket{
      \tikz[baseline=-0.5ex, scale=\tikzscale]{
        \draw (1.0ex, -1.0ex) -- (1.0ex, 1.0ex);
        \draw (0, 0) -- (2ex, 0);
        \fill[black, draw=black] (0,0) circle (0.4ex);
        \fill[black, draw=black] (2ex,0) circle (0.4ex);
        \fill[white, draw=black] (1ex,1ex) circle (0.4ex);
        \fill[white, draw=black] (1ex,-1ex) circle (0.4ex);
      }
    }\!
    \bra{
      \tikz[baseline=-0.5ex, scale=\tikzscale]{
        \draw (1.0ex, -1.0ex) -- (1.0ex, 1.0ex);
        \draw (0, 0) -- (2ex, 0);
        \fill[white, draw=black] (0,0) circle (0.4ex);
        \fill[white, draw=black] (2ex,0) circle (0.4ex);
        \fill[black, draw=black] (1ex,1ex) circle (0.4ex);
        \fill[black, draw=black] (1ex,-1ex) circle (0.4ex);
      }
    } +\  \text{h.c.} 
$, which energetically selects the pVBS ground state via an order-by-disorder mechanism~\cite{CastroNeto2006, Hermele2004, Henry2014}. Notably, the ice-rule constraint is a subset of the $B_p = 1$ constraint, which also allows ``all-in/all-out'' configurations (see Fig.~\ref{fig:models}). However, if one selects, as a star operator of 2-term model,
\[
A_s =
  \ket{
      \tikz[baseline=-0.5ex, scale=\tikzscale]{
        \draw (1.0ex, -1.0ex) -- (1.0ex, 1.0ex);
        \draw (0, 0) -- (2ex, 0);
        \fill[black, draw=black] (0,0) circle (0.4ex);
        \fill[black, draw=black] (2ex,0) circle (0.4ex);
        \fill[white, draw=black] (1ex,1ex) circle (0.4ex);
        \fill[white, draw=black] (1ex,-1ex) circle (0.4ex);
      }
    }\!
    \bra{
      \tikz[baseline=-0.5ex, scale=\tikzscale]{
        \draw (1.0ex, -1.0ex) -- (1.0ex, 1.0ex);
        \draw (0, 0) -- (2ex, 0);
        \fill[white, draw=black] (0,0) circle (0.4ex);
        \fill[white, draw=black] (2ex,0) circle (0.4ex);
        \fill[black, draw=black] (1ex,1ex) circle (0.4ex);
        \fill[black, draw=black] (1ex,-1ex) circle (0.4ex);
      }
    } +\  \text{h.c.}
\] the operator does not convert two-in-two-out tetrahedras into all-in/all-out ones. As a result, configurations containing all-in/all-out tetrahedra remain dynamically disconnected from the TFIM subspace. Figure~\ref{fig:TFIM}(b) shows the low-lying spectra of the symmetry-deformed TC model and the TFIM (evaluated using only the leading-order term), showing that the lowest energy levels are isospectral.  Additional states in the symmetry-deformed TC model arise from configurations involving all-in/all-out tetrahedra. An equivalent model can also be realized in the odd gauge $(B_p = -1)$ sector by choosing terms that favor plaquette configurations whose uniform tiling satisfies the $B_p$ constraint (see Fig.~\ref{fig:models}).

\begin{figure}[t]
  \centering
  \includegraphics[width=1.0\linewidth]{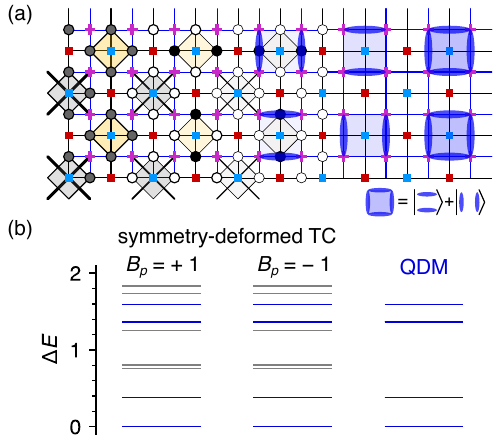}
  \caption{{\bf Equivalence between QDM and symmetry-deformed TC models.}
  (a) Schematics of a spin/dimer configuration for symmetry-deformed TC/QDM.
  (b) Energy spectra of the QDM and two variants of the QDM-type two-term symmetry-deformed TC model with uniform $B_p = +1$ and $B_p = -1$ constraints, obtained via ED for a system of size $N = 32$. While all three models share the same low-energy spectra (indicated by the blue lines), the symmetry-deformed TC models also exhibit additional states that are dynamically disconnected from the QDM subspace.}
  \label{fig:isospcetrum}
\end{figure}

The equivalence between the QDM and the symmetry-deformed TC model is most transparent in the $B_p = -1$ sector with 
\[
A_s =
\ket{
  \tikz[baseline=-0.5ex, scale=\tikzscale]{
    \draw (1.0ex, -1.0ex) -- (1.0ex, 1.0ex);
    \draw (0, 0) -- (2ex, 0);
    \fill[black, draw=black] (0,0) circle (0.4ex);
    \fill[black, draw=black] (2ex,0) circle (0.4ex);
    \fill[white, draw=black] (1ex,1ex) circle (0.4ex);
    \fill[white, draw=black] (1ex,-1ex) circle (0.4ex);
  }
}\!
\bra{
  \tikz[baseline=-0.5ex, scale=\tikzscale]{
    \draw (1.0ex, -1.0ex) -- (1.0ex, 1.0ex);
    \draw (0, 0) -- (2ex, 0);
    \fill[white, draw=black] (0,0) circle (0.4ex);
    \fill[white, draw=black] (2ex,0) circle (0.4ex);
    \fill[black, draw=black] (1ex,1ex) circle (0.4ex);
    \fill[black, draw=black] (1ex,-1ex) circle (0.4ex);
  }
} +\  \text{h.c.}
\]
where the presence or absence of a dimer can be identified with an up or down spin, respectively, in the spin representation defined on the dual lattice [Fig.~\ref{fig:isospcetrum}(a)]. The dimer constraint---requiring exactly one dimer per vertex---ensures that the system lies within the $B_p = -1$ sector. As in the TFIM case, the $B_p = -1$ constraint in the QDM mapping can also be satisfied by configurations with three dimers attached to a single vertex. However, such states are dynamically disconnected from the QDM subspace (with exactly one dimer per vertex), as the star term (kinetic term) preserves the number of dimers per vertex.
Consequently, as long as the ground state of the symmetry-deformed TC model lies within the QDM subspace, the two models exhibit identical low-energy spectra---up to additional exact copies arising from the disconnected subspace in which every vertex hosts three dimers. Figure~\ref{fig:isospcetrum}(b) shows the low-lying spectra of the QDM and a symmetry-deformed TC model with $B_p = -1$. The two spectra match exactly within the QDM subspace, while the symmetry-deformed TC model also features extra states corresponding to configurations with vertices occupied by three dimers.

For the $B_p = +1$ constraint, the connection between the QDM and the symmetry-deformed TC model is less direct. However, one can still establish an exact mapping by redefining the dimer-spin correspondence. For example, in the model with $A_s = 
\ket{
  \tikz[baseline=-0.5ex, scale=\tikzscale]{
    \draw (1.0ex, -1.0ex) -- (1.0ex, 1.0ex);
    \draw (0, 0) -- (2ex, 0);
    \fill[white, draw=black] (0,0) circle (0.4ex);
    \fill[white, draw=black] (2ex,0) circle (0.4ex);
    \fill[white, draw=black] (1ex,1ex) circle (0.4ex);
    \fill[black, draw=black] (1ex,-1ex) circle (0.4ex);
  }
}\!
\bra{
  \tikz[baseline=-0.5ex, scale=\tikzscale]{
    \draw (1.0ex, -1.0ex) -- (1.0ex, 1.0ex);
    \draw (0, 0) -- (2ex, 0);
    \fill[black, draw=black] (0,0) circle (0.4ex);
    \fill[black, draw=black] (2ex,0) circle (0.4ex);
    \fill[black, draw=black] (1ex,1ex) circle (0.4ex);
    \fill[white, draw=black] (1ex,-1ex) circle (0.4ex);
  }
}
+\  \text{h.c.}$, the spin-down/up state on the upper site of an $s_1$ vertex may be identified with the presence/absence of a dimer, while for all other sites, the usual correspondence of spin-down/up to absence/presence of a dimer applies. The ED spectrum confirms the exact equivalence between the spectra of different QDM-type two-term models.

\section{Quantum Dimer Model}
\label{sec:QDM}
\begin{figure}[t]
    \centering
    \includegraphics[width=1.0\linewidth]{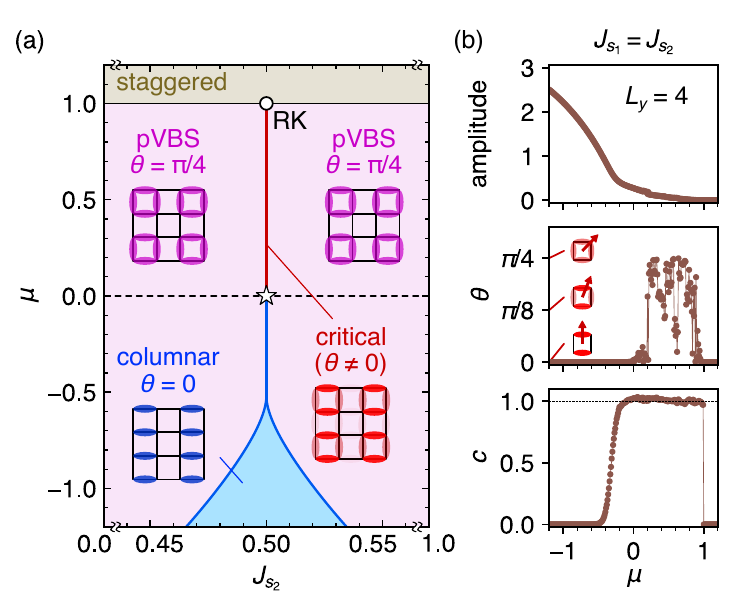}
    \caption{{\bf Quantum dimer model.}
    (a) Phase diagram as a function of the chemical potential $\mu$ and the star term strength $J_{s_2}$.
    (b) $\mu$-dependence of physical quantities extracted from iDMRG at $J_{s_1} = J_{s_2}$. From top to bottom: amplitude of the order parameter, angle $\theta$, and central charge $c$ (see text for details). The angle $\theta = 0$ and $\theta = \pi/4$ correspond to the columnar and pVBS phases, respectively, while intermediate values $0 < \theta < \pi/4$ indicate a mixed phase.}
    \label{fig:QDM}
\end{figure}

Having established the equivalence between the QDM and the symmetry-deformed TC model, we now turn our attention to the QDM itself. The Hamiltonian of the QDM on the square lattice is given by \cite{Moessner2011}
\begin{equation}
    H_{\text{QDM}} = \sum_{p'}
    \left[-t \left(\ket{
    \tikz[baseline=-0.5ex, scale=\tikzscale]{
        \fill (0,1ex) circle (0.5ex);
        \fill (0,-1ex) circle (0.5ex);
        \fill (2ex,1ex) circle (0.5ex);
        \fill (2ex,-1ex) circle (0.5ex);
        \draw[thick] (0,1ex) -- (2ex,1ex);
        \draw[thick] (0,-1ex) -- (2ex,-1ex);
    }
    }
    \bra{
    \tikz[baseline=-0.5ex, scale=\tikzscale]{
        \fill (0,1ex) circle (0.5ex);
        \fill (0,-1ex) circle (0.5ex);
        \fill (2ex,1ex) circle (0.5ex);
        \fill (2ex,-1ex) circle (0.5ex);
        \draw[thick] (0,1ex) -- (0,-1ex);
        \draw[thick] (2ex,1ex) -- (2ex,-1ex);
    }
    } + \text{h.c.}\right)
    + \mu \left(\ket{
    \tikz[baseline=-0.5ex, scale=\tikzscale]{
        \fill (0,1ex) circle (0.5ex);
        \fill (0,-1ex) circle (0.5ex);
        \fill (2ex,1ex) circle (0.5ex);
        \fill (2ex,-1ex) circle (0.5ex);
        \draw[thick] (0,1ex) -- (2ex,1ex);
        \draw[thick] (0,-1ex) -- (2ex,-1ex);
    }
    }
    \bra{
    \tikz[baseline=-0.5ex, scale=\tikzscale]{
        \fill (0,1ex) circle (0.5ex);
        \fill (0,-1ex) circle (0.5ex);
        \fill (2ex,1ex) circle (0.5ex);
        \fill (2ex,-1ex) circle (0.5ex);
        \draw[thick] (0,1ex) -- (2ex,1ex);
        \draw[thick] (0,-1ex) -- (2ex,-1ex);
    }
    } +
    \ket{
    \tikz[baseline=-0.5ex, scale=\tikzscale]{
        \fill (0,1ex) circle (0.5ex);
        \fill (0,-1ex) circle (0.5ex);
        \fill (2ex,1ex) circle (0.5ex);
        \fill (2ex,-1ex) circle (0.5ex);
        \draw[thick] (0,1ex) -- (0,-1ex);
        \draw[thick] (2ex,1ex) -- (2ex,-1ex);
    }
    }
    \bra{
    \tikz[baseline=-0.5ex, scale=\tikzscale]{
        \fill (0,1ex) circle (0.5ex);
        \fill (0,-1ex) circle (0.5ex);
        \fill (2ex,1ex) circle (0.5ex);
        \fill (2ex,-1ex) circle (0.5ex);
        \draw[thick] (0,1ex) -- (0,-1ex);
        \draw[thick] (2ex,1ex) -- (2ex,-1ex);
    }
    }
    \right)
    \right],
\end{equation}
with $t$ and $\mu$ parametrizing the relative strength of dimer resonance and chemical potential terms, respectively. 
The case $\mu = 0$ corresponds to the 2-term model indicated by the star symbol in the phase diagram of Fig.~\ref{fig:QDM}(a).

Despite extensive studies of the QDM~\cite{Syljuaasen2005, Syljuaasen2006, Ralko2008, Banerjee2014, Banerjee2016, Yan2021}, the ground-state phase diagram for $\mu < 1$ remains unsettled. In particular, it is still unclear whether the columnar phase---characterized by translational symmetry breaking along one of the principal lattice directions---persists all the way to the Rokhsar-Kivelson (RK) point ($\mu = 1$). An alternative scenario is the presence of a gapped mixed (or pVBS) phase before reaching the RK point \cite{Ralko2008, Yan2021}. Like the pVBS, the mixed phase breaks translational symmetry in two directions, and in addition---like the columnar phase---breaks $C_4$ symmetry. With $\theta$ quantifying the strength of $C_4$ symmetry breaking, the mixed phase can be characterized by an angle $0 < \theta < \pi/4$. Identifying phases can be challenging, as small (large) values of $\theta$ can blur the distinction between mixed and columnar (pVBS) phases \cite{Yan2021}, and any finite-temperature numerics may be subject to undesired thermal entropic bias~\cite{Henley1989}.

Notably, the sublattice modulation technique employed in the analysis of symmetry-deformed TC models above provides valuable insight also into this problem (Fig.~\ref{fig:QDM}). Unlike the ambiguous nature of the $J_{s_1} = J_{s_2}$ point, i.e., the original QDM, a broad region of the extended phase diagram at $J_{s_1} \neq J_{s_2}$ clearly supports pVBS order, smoothly connected to the exactly solvable limit $J_{s_2} = 0$. In contrast, the columnar phase in the QDM limit remains stable under sublattice modulation only at large negative $\mu$. The energy gap between the columnar and pVBS phases can be roughly estimated by the critical value $(J_{s_2})_\text{c}$ for a given $\mu$, and numerics indicate that this gap decreases with increasing $\mu$. For instance, at $\mu = -0.25$, $|(J_{s_2})_\text{c} - 0.5|$ is as small as $5.0 \times 10^{-6}$, and for $\mu > 0$, the system appears practically critical. 

To identify a possible mixed phase, we define an order parameter amplitude as $\sqrt{\sigma^z(\pi, \pi)^2 + \sigma^z(\pi, -\pi)^2}$,  with $\sigma^z(\vec{k}) \equiv \sum_i \text{e}^{-\text{i} \vec{k} \cdot \vec{r}_i} \braket{\sigma^z_i}$, and the angle $\theta$ as $\theta = \arctan\left[\sigma^z(\pi, \pi)/\sigma^z(\pi, -\pi)\right]$, assuming $\sigma^z(\pi, \pi) \leq \sigma^z(\pi, -\pi)$. 
At $J_{s_1} = J_{s_2}$ and $\mu \gtrsim 0.0$, iDMRG struggles to consistently resolve an optimal value of $\theta$ [Fig.~\ref{fig:QDM}(b)]---a wide range of $\theta \in [0, \pi/4]$ yields nearly degenerate energies, suggesting an emergent SO(2) symmetry with no clear energetic preference for a particular angle. Our results resemble the approximate SO(2) symmetry reported in Ref.~\cite{Banerjee2014, Banerjee2016}, except that, in our case, no small but finite gap is detected within numerical accuracy. This absence of a resolvable gap is corroborated by the instability of the $J_{s_1} = J_{s_2}$ phase under minimal sublattice modulation. The critical nature is also indicated by a central charge estimate $c \approx 1.0$ [Fig.~\ref{fig:QDM}(b)], calculated from the entanglement entropy, consistent with the presence of a gapless mode~\cite{Tagliacozzo2008, Pollmann2009}. Whether, however, this criticality is genuine or merely a proximity effect arising from second-order transitions in the direct (and numerically not resolvable) vicinity remains an open question for future investigation.
To summarize, by extending the QDM phase space, we identify competing stability regions of pVBS and columnar phases, and find that the QDM, as a symmetry-deformed TC, exhibits (within our numerical accuracy) quantum critical behavior with an emergent SO(2) symmetry.

\section{Discussion}
Unlike symmetry-based gauging and ungauging procedures ---which typically relate fixed-point Hamiltonians---our symmetry deformation can yield models with distinct characteristics that reflect microscopic details, such as subsystem symmetries or the specific type of lattice geometry (e.g., square, triangular, or honeycomb). Although this complexity introduces analytical challenges and may necessitate numerical methods, it is precisely because of this extra level of specificity that our term-dropping protocol may also lead to various emergent phases that simple gauging or ungauging approaches cannot capture---making it a compelling avenue for exploring new routes to complex quantum states.

For the square lattice $\mathbb{Z}_2$ TO, starting from the TC, we have shown that introducing a global $U(1)$ symmetry leads to a breakdown of the underlying gauge structure, resulting in a pVBS phase. This transition can be understood in terms of $e$-anyon condensation, analogous to the field-induced breakdown of TO in the undeformed TC. The same symmetry-deformation approach leads to a variety of other models, including the $XY$TC and four-term models that exhibit subsystem symmetries---remnants of the original 1-form symmetry of the TC.

As extreme cases within the family of symmetry-deformed TC models, one ends up with instances of the TFIM on the checkerboard lattice and the QDM. Employing the same sublattice modulation technique,
which has proven to be a powerful tool for exploring the phase diagram of symmetry-deformed TCs, one can add a new perspective on these models. For the QDM, this has allowed us to identify stable regions of columnar ordering and argue for an emergent SO(2) symmetry in the original QDM.

An interesting future extension of this work is to explore TOs with larger gauge groups than the $\mathbb{Z}_2$ TO considered here. One such example would be $\mathbb{Z}_4$, where intermediate TO phases, including SET, naturally emerge through partial ungauging. It will be an interesting question to explore whether the symmetry-deformation approach can also realize such intermediate topological phases, potentially revealing a broader landscape of novel quantum orders.\\


{\it Data availability}.-- 
The numerical data shown in the figures and the data for sweeping the phase diagram is available on Zenodo~\cite{zenodo}. Our iDMRG simulations employed the publicly available Python package \texttt{TenPy}~\cite{Tenpy2024}.\\

{\it Note added}.-- While finalizing this work, we became aware of a related independent study~\cite{Vieweg2025}, which is expected to appear on arXiv concurrently.


\acknowledgments
We thank Bo Han, Zemin Huang, Zohar Nussinov, and especially Ciar\'an Hickey for discussions.
We gratefully acknowledge partial funding from the Deutsche Forschungsgemeinschaft (DFG, German Research Foundation)
under Germany's Excellence Strategy---Cluster of Excellence Matter and Light for Quantum Computing (ML4Q) EXC 2004/1 -- 390534769,
within the CRC network TR 183 (Project Grant No.~277101999) as part of subproject B01,
as well as within SFB 1238 (Project Grant No.~277146847) as part of project C03.
The numerical simulations were performed on the JUWELS cluster at the Forschungszentrum Juelich, the Noctua2 cluster at PC$^2$ in Paderborn, 
and the RAMSES cluster at RRZK Cologne.

\bibliography{ref}


\appendix

\section{Numerical simulations}
\label{app:numerics}

\begin{figure}[t]
  \centering
  \includegraphics[width=1.0\linewidth]{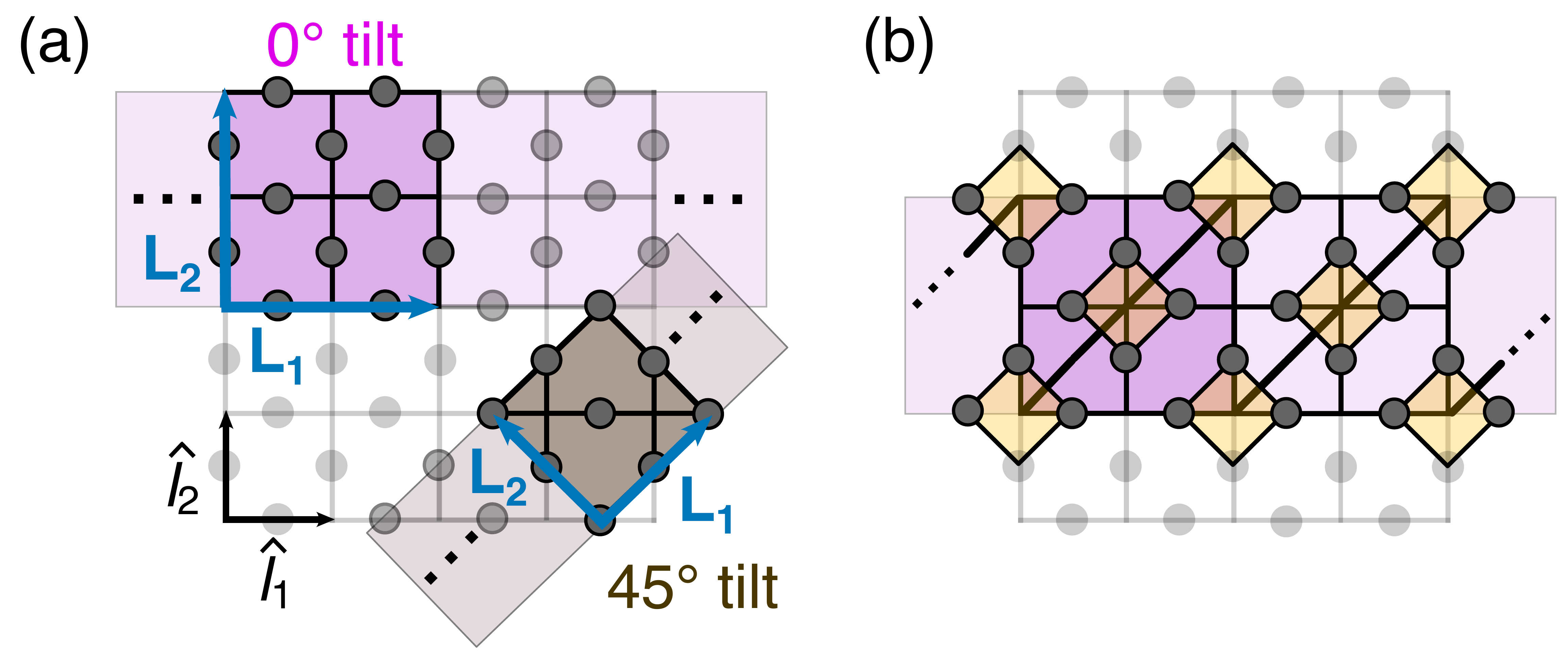}
   \caption{{\bf Compactification schemes.} 
   	(a) \(0^{\circ}\) tilt compactification with $L_x = L_y = 2$ and \(45^{\circ}\) tilt compactification with $L_x = L_y = 1$. 
		The numerical unit cell is repeated infinitely along the \(\vec{L}_1\) direction, while the \(\vec{L}_2\) direction is finite and periodic.
    	(b) \(0^{\circ}\) tilt compactification used for the $U(1)$TC, 
		where the diagonal direction extends indefinitely along the infinite direction of the cylinder.
   }
   \label{fig:compactification scheme}
\end{figure}

Here, we provide a detailed description of the numerical setup used in our iDMRG simulations~\cite{Schollwoeck2011}. We consider a square lattice spanned by the two principal lattice vectors $\hat{l}_1 = (1, 0)$ and $\hat{l}_2 = (0, 1)$, with spin-1/2 degrees of freedom residing on the links. The numerical unit cell is defined by two orthogonal vectors $\vec{L}_1$ and $\vec{L}_2$:
\begin{equation*}
    \vec{L}_1 = L_x(a\hat{l}_1 + b\hat{l}_2), \quad \vec{L}_2 = L_y(-b\hat{l}_1 + a\hat{l}_2),
\end{equation*}
where integers $(a, b)$ specify the angle of $\vec{L}_1$ with respect to $\hat{l}_1$. For instance, $(a, b) = (1, 0)$ corresponds to 0$^\circ$ compactification and $(1, 1)$ to 45$^\circ$ compactification [Fig.~\ref{fig:compactification scheme}(a)]. We use an infinite cylinder geometry, with $\vec{L}_1$ as the infinite direction (repeated blocks of size $L_x$) and $\vec{L}_2$ as the finite periodic direction of length $L_y$.

\subsubsection*{\(U(1)\) toric code simulations}

In the exactly solvable limit $J_{s_2} = 0$, the $U(1)$ toric code exhibits a subextensive ground-state degeneracy, and finite $J_{s_2}$ acts as a perturbation on this classical degeneracy. In iDMRG simulations, global fluctuations---induced by operators $\prod_{s_2 \in C_{\text{(anti-)diag}}} A_{s_2}$ along diagonal ($\hat{l}_1 + \hat{l}_2$) and anti-diagonal ($\hat{l}_1 - \hat{l}_2$) directions---can complicate the analysis of the ground state. However, as we explain below, such fluctuations can be suppressed by choosing a suitable compactification scheme.

With 45$^\circ$ compactification, these global fluctuations can still be induced by repeated action along the finite periodic direction $\vec{L}_2$. This causes a typical finite-size effect seen, for example, in exact diagonalization, where the finite system size forces the ground state to be a symmetric combination of all configurations related by global or subsystem symmetry operations. In contrast, with 0$^\circ$ compactification, both diagonal and anti-diagonal directions extend indefinitely along the infinite direction of the cylinder [Fig.~\ref{fig:compactification scheme}(b)], entirely suppressing global fluctuations. Consequently, only local fluctuations remain, simplifying the analysis of the ground state. The same principle applies to the $XY$TC model, where, in the absence of global fluctuations, spontaneous symmetry-breaking states can be observed by taking the subsystem symmetry direction to be infinite.

Most of the results presented in the main text were obtained using the $0^\circ$ compactification scheme, with numerical unit-cell dimensions $L_x = L_y = 4$. The bond dimension $\chi$ was increased up to $800$, ensuring that the truncation error remained below $6 \times 10^{-7}$. To accurately estimate the prefactor $\beta$ in the finite-size energy splitting (see main text), we also studied larger circumferences ($L_y = 6$ and $8$), avoiding any potential overestimation of the energy gain from the fourth-order perturbation. (For $L_y = 4$, repeated actions of the $A_{s_2}$ operators along the finite periodic $L_y$ direction can induce additional local fluctuations.)

\subsubsection*{Quantum dimer model simulations}

The possible ground states of the quantum dimer model discussed in the literature include the columnar, pVBS, and mixed phases~\cite{Syljuaasen2005, Syljuaasen2006, Ralko2008, Banerjee2014, Banerjee2016, Yan2021}. It is thus crucial to ensure that the choice of geometry in iDMRG simulations does not inadvertently introduce anisotropies that could bias the numerical results. In particular, special caution is required when studying the columnar and mixed phases, which involves the $C_4$ rotational symmetry breaking of the square lattice.

For instance, with $0^\circ$ compactification, the simulations preferentially select the columnar phase that breaks symmetry along the infinite $\hat{l}_1$ direction, rather than the finite periodic $\hat{l}_2$ direction. This preference arises because, in order to stabilize translational symmetry breaking along a finite periodic direction, the system would need to form a superposition of the two possible symmetry-breaking patterns rather than realizing a true symmetry-breaking configuration. Such superpositions typically lead to higher entanglement entropy and are therefore numerically disfavored in iDMRG.

In contrast, the $45^\circ$ compactification, which preserves $C_4$ symmetry, allows for symmetry breaking along both lattice directions. Numerically, we confirmed that in this case, the simulation indeed selects one of the four possible symmetry-breaking patterns, depending on the initial state. Therefore, for the quantum dimer model, we typically use the $45^\circ$ compactification scheme.

The numerical unit-cell dimensions $L_x = L_y = 4$ and a bond dimension $\chi$ up to $1000$ are used. We tested both pVBS and columnar initial states in our simulations, each obtained from parameter regimes deep inside their respective phases. In the parameter regime where the two phases compete, i.e., $0 < \mu < 1$ with $J_{s_1} = J_{s_2}$, where we observed (nearly) critical behavior, the choice of initial state can lead to differences in the monitored physical quantities, as each result tends to be biased towards the respective phase. However, no apparent energy differences larger than $2 \times 10^{-8}$ were found in the converged iDMRG results, further supporting the critical nature of the system. The order parameters in the main text, are constructed from $\sigma^z(\vec{k}) \equiv \sum_i \text{e}^{-\text{i} \vec{k} \cdot \vec{r}_i} \braket{\sigma^z_i}$, where $\vec{r}_i$ is defined using the $\vec{L}_1$ and $\vec{L}_2$ directions as a basis, with the spacing between sites along these directions set to one.

\end{document}